\def\IEEEFORMAT{}

\ifdefined\ACMFORMAT
  \documentclass[acmsmall]{acmart}

\usepackage{balance}

\usepackage[ruled,vlined]{algorithm2e}
\usepackage{array}
\usepackage{multirow}
\usepackage{subfigure}
\usepackage{tikz}
\usepackage{pgfplots}
\usepackage{enumitem}
\usepackage{colortbl}
\usepackage{pdfpages}

\pgfplotsset{compat=1.18}

\newtheorem{theorem}{Theorem}
\newtheorem{lemma}{Lemma}
\newtheorem{definition}{Definition}

\acmJournal{TODAES}
\acmVolume{1}
\acmNumber{1}
\acmArticle{1}
\acmYear{2024}
\acmMonth{1}
\acmDOI{10.1145/1122445.1122456}

\setcopyright{acmcopyright}
\copyrightyear{2024}
\acmPrice{15.00}

\newcommand{\acmtitle}[1]{\title{#1}}
\newcommand{\acmauthor}[2]{\author{#1}\affiliation{\institution{#2}\city{City}\country{Country}}}

\newcommand{\acmccs}[1]{%
  \ccsdesc[500]{Computer systems organization~Architectures}%
  \ccsdesc[300]{Hardware~Application-specific VLSI designs}%
  \ccsdesc[300]{Hardware~Functional verification}%
  \ccsdesc[100]{Hardware~Electronic design automation}%
}

\newcommand{\acmabstract}[1]{%
  #1%
  \maketitle%
}

\else
  \def\IEEEFORMAT{}%
  \documentclass[journal]{IEEEtran}

\usepackage{graphicx}
\usepackage{xpatch}

\usepackage[utf8]{inputenc}
\usepackage[T1]{fontenc}
\usepackage{amsmath,amsfonts,amssymb}
\usepackage{amsthm}
\usepackage{cite}
\usepackage{url}
\usepackage{balance}

\usepackage{arydshln}
\usepackage{centernot}
\usepackage{pbox}
\usepackage{bm}
\usepackage{comment}
\usepackage{subfigure}
\usepackage{epstopdf}
\usepackage{multirow}
\usepackage[table]{xcolor}
\usepackage{tabu}
\usepackage{mdframed}
\usepackage{fancyvrb}

\usepackage{algorithm}           
\usepackage{algorithmicx}        
\usepackage[noend]{algpseudocode} 
\usepackage{siunitx}
\usepackage{paralist}
\usepackage{stmaryrd}
\usepackage{tikz}
\usepackage[referable]{threeparttablex}
\usepackage{tabularx}
\usepackage{booktabs}
\usepackage{array}
\usepackage{ragged2e}
\usepackage{fancyhdr}
\usepackage{float}
\usepackage{xspace}
\usepackage{pifont}
\usepackage{calc}
\usepackage{caption}
\usepackage{enumitem}
\usepackage{lettrine}  
\usepackage{pgfplots}
\usepgfplotslibrary{statistics}
\usepackage[hidelinks]{hyperref}
\usepackage{orcidlink}
\usepackage{cleveref}
\usepackage{pdfpages}

\usetikzlibrary{shapes,arrows, positioning}
\usetikzlibrary{decorations.pathmorphing}

\pgfplotsset{compat=1.18}

\definecolor{princetonorange}{RGB}{255,143,0}

\newcolumntype{Y}{>{\RaggedRight\arraybackslash}X}

\newcommand{\Init}{\mathit{Init}}

\newcommand{\ieeetitle}[1]{\title{#1}}
\newcommand{\ieeeauthor}[1]{\author{#1}}

\newcommand{\ieeekeywords}[1]{%
  \begin{IEEEkeywords}
  #1
  \end{IEEEkeywords}
  \IEEEpeerreviewmaketitle
}

\fi

\newcommand{\vkw}[1]{\textcolor{green!45!black}{#1}}
\newcommand{\vdecl}[1]{\textcolor{red!70!black}{#1}}
\newcommand{\vtype}[1]{\textcolor{purple!70!black}{#1}}
\newcommand{\vlit}[1]{\textcolor{blue!70!black}{#1}}
\newcommand{\vcmt}[1]{\textcolor{gray!75!black}{#1}}

\newcommand{\hongce}[1]{\textcolor{red}{[\textbf{hongce}: #1]}}

\def\BibTeX{{\rm B\kern-.05em{\sc i\kern-.025em b}\kern-.08em
    T\kern-.1667em\lower.7ex\hbox{E}\kern-.125emX}}

\newlist{myitemize}{itemize}{1}
\setlist[myitemize]{
  label=\textbullet, 
}

\usepackage{etoolbox}
\makeatletter
\patchcmd{\@makecaption}
  {\scshape}
  {}
  {}
  {}
\makeatother

\begin{document}

\ifdefined\IEEEFORMAT
  \ieeetitle{Enhancing Word-Level Property Directed Reachability with LLM-Driven Semantic Guidance}
  \ieeeauthor{%
    Guangyu~Hu~\orcidlink{0000-0001-5077-8361}\textsuperscript{*},
    Mingkai~Miao~\orcidlink{0009-0001-0434-5933}\textsuperscript{*}, \IEEEmembership{Student Member,~IEEE},
    Zhiyuan~Yan~\orcidlink{0000-0003-3857-6649},
    \IEEEmembership{Member,~IEEE},
    \mbox{Xiaofeng~Zhou~\orcidlink{0000-0001-5878-3683},~\IEEEmembership{Graduate Student Member,~IEEE}},
    \mbox{Wei~Zhang~\orcidlink{0000-0002-7622-6714}},~\IEEEmembership{Fellow,~IEEE},
    and Hongce~Zhang~\orcidlink{0000-0003-4001-264X}\textsuperscript{\textdagger},~\IEEEmembership{Member,~IEEE}%
    \thanks{\textsuperscript{*}Guangyu Hu and Mingkai Miao contributed equally to this work.}%
    \thanks{Guangyu Hu and Xiaofeng Zhou are with The Hong Kong University of Science and Technology,
    Hong Kong SAR, China
    (e-mail: ghuae@connect.ust.hk; xzhoubu@connect.ust.hk).}%
    \thanks{Mingkai Miao, Zhiyuan Yan, and Hongce Zhang are with the Microelectronics Thrust,
    Function Hub, Hong Kong University of Science and Technology (Guangzhou),
    Guangzhou 511458, China
    (e-mail:~mmiao815@connect.hkust-gz.edu.cn;~zyan760@connect.hkust-gz.edu.cn;~hongcezh@hkust-gz.edu.cn).}%
    \thanks{Wei Zhang is with the Department of Electronic and Computer Engineering,
    The Hong Kong University of Science and Technology, Hong Kong, China
    (e-mail: eeweiz@ust.hk).}%
    \thanks{\textsuperscript{\textdagger}Hongce Zhang is the corresponding author.}}
  \maketitle
  \markboth{}{}
  \begin{abstract}
Property Directed Reachability (PDR) is a prominent algorithm for hardware formal verification. However, bit-level PDR often struggles with datapath-heavy designs because bit-blasting obscures high-level semantics. While word-level PDR addresses this by reasoning over bit-vector and array theories, its performance remains bottlenecked by discovering proof-relevant word-level relations. We propose \texttt{LLM4PDR}, a framework leveraging Large Language Models (LLMs) to guide word-level PDR search through three mechanisms: (1) \emph{Predicate Generation}, extracting state relationships as candidate predicates during inductive generalization; (2) \emph{Clause Generation}, producing candidate frame lemmas to accelerate convergence after formal validation; and (3) \emph{Assertion Generation}, synthesizing helper assertions that strengthen the target property under counterexample-guided refinement. We implement \texttt{LLM4PDR} in the Pono model checker and evaluate it on arithmetic micro-benchmarks, HLS-generated pipelines, open-source RTL components, and hardware model checking competition (HWMCC) instances. Results show LLM-generated guidance improves both solved instances and runtime on datapath-heavy and control-plus-datapath designs. The strongest configuration solves 28 of 33 arithmetic benchmarks, compared to 13 for vanilla Pono and 11 for AVR. On HLS pipelines and open-source RTL, different modes provide complementary speedups: clause guidance is effective for deep pipelines and predicate guidance for bus and memory-controller designs. On HWMCC benchmarks, benefits are instance-dependent, demonstrating notable speedups and timeout avoidance on hard cases. These results suggest that LLM-generated, verifier-checked semantic hints can serve as a practical complement to conventional word-level PDR.
\end{abstract}
  \ieeekeywords{Hardware formal verification, property directed reachability, word-level model checking, large language models, invariant generation, inductive generalization.}
\else\ifdefined\ACMFORMAT
  \acmtitle{Enhancing Word-Level Property Directed Reachability with LLM-Driven Semantic Guidance}
  \acmauthor{Author 1}{University Name}
  \acmauthor{Author 2}{University Name}
  \acmauthor{Author 3}{University Name}



  \acmabstract{\begin{abstract}
Property Directed Reachability (PDR) is a prominent algorithm for hardware formal verification. However, bit-level PDR often struggles with datapath-heavy designs because bit-blasting obscures high-level semantics. While word-level PDR addresses this by reasoning over bit-vector and array theories, its performance remains bottlenecked by discovering proof-relevant word-level relations. We propose \texttt{LLM4PDR}, a framework leveraging Large Language Models (LLMs) to guide word-level PDR search through three mechanisms: (1) \emph{Predicate Generation}, extracting state relationships as candidate predicates during inductive generalization; (2) \emph{Clause Generation}, producing candidate frame lemmas to accelerate convergence after formal validation; and (3) \emph{Assertion Generation}, synthesizing helper assertions that strengthen the target property under counterexample-guided refinement. We implement \texttt{LLM4PDR} in the Pono model checker and evaluate it on arithmetic micro-benchmarks, HLS-generated pipelines, open-source RTL components, and hardware model checking competition (HWMCC) instances. Results show LLM-generated guidance improves both solved instances and runtime on datapath-heavy and control-plus-datapath designs. The strongest configuration solves 28 of 33 arithmetic benchmarks, compared to 13 for vanilla Pono and 11 for AVR. On HLS pipelines and open-source RTL, different modes provide complementary speedups: clause guidance is effective for deep pipelines and predicate guidance for bus and memory-controller designs. On HWMCC benchmarks, benefits are instance-dependent, demonstrating notable speedups and timeout avoidance on hard cases. These results suggest that LLM-generated, verifier-checked semantic hints can serve as a practical complement to conventional word-level PDR.
\end{abstract}}
\else
  \ieeetitle{[Your Paper Title Here]: A General Framework for [Your Research Area]}
  \ieeeauthor{Author 1, Author 2, and Author 3}
  \maketitle
  \begin{abstract}
Property Directed Reachability (PDR) is a prominent algorithm for hardware formal verification. However, bit-level PDR often struggles with datapath-heavy designs because bit-blasting obscures high-level semantics. While word-level PDR addresses this by reasoning over bit-vector and array theories, its performance remains bottlenecked by discovering proof-relevant word-level relations. We propose \texttt{LLM4PDR}, a framework leveraging Large Language Models (LLMs) to guide word-level PDR search through three mechanisms: (1) \emph{Predicate Generation}, extracting state relationships as candidate predicates during inductive generalization; (2) \emph{Clause Generation}, producing candidate frame lemmas to accelerate convergence after formal validation; and (3) \emph{Assertion Generation}, synthesizing helper assertions that strengthen the target property under counterexample-guided refinement. We implement \texttt{LLM4PDR} in the Pono model checker and evaluate it on arithmetic micro-benchmarks, HLS-generated pipelines, open-source RTL components, and hardware model checking competition (HWMCC) instances. Results show LLM-generated guidance improves both solved instances and runtime on datapath-heavy and control-plus-datapath designs. The strongest configuration solves 28 of 33 arithmetic benchmarks, compared to 13 for vanilla Pono and 11 for AVR. On HLS pipelines and open-source RTL, different modes provide complementary speedups: clause guidance is effective for deep pipelines and predicate guidance for bus and memory-controller designs. On HWMCC benchmarks, benefits are instance-dependent, demonstrating notable speedups and timeout avoidance on hard cases. These results suggest that LLM-generated, verifier-checked semantic hints can serve as a practical complement to conventional word-level PDR.
\end{abstract}
\fi\fi

\section{Introduction}
\label{sec:intro}

The ever-increasing complexity of modern hardware designs poses significant challenges to functional verification. Formal verification, particularly automated model checking, has emerged as a critical technology for ensuring the correctness of these systems by exhaustively exploring possible behaviors. Among model checking techniques, Property Directed Reachability (PDR), also known as IC3~\cite{bradley2011ic3}, is a leading algorithm for hardware formal verification. Traditional bit-level PDR is highly effective for many control-centric designs, but modern designs often combine control logic with wide datapaths, arithmetic operations, and on-chip memories. Bit-blasting these word-level structures can break compact arithmetic and relational facts into many low-level Boolean constraints, obscuring the design intent that would otherwise guide the proof search.

\begin{figure}[t]
    \centering
    \includegraphics[width=\columnwidth]{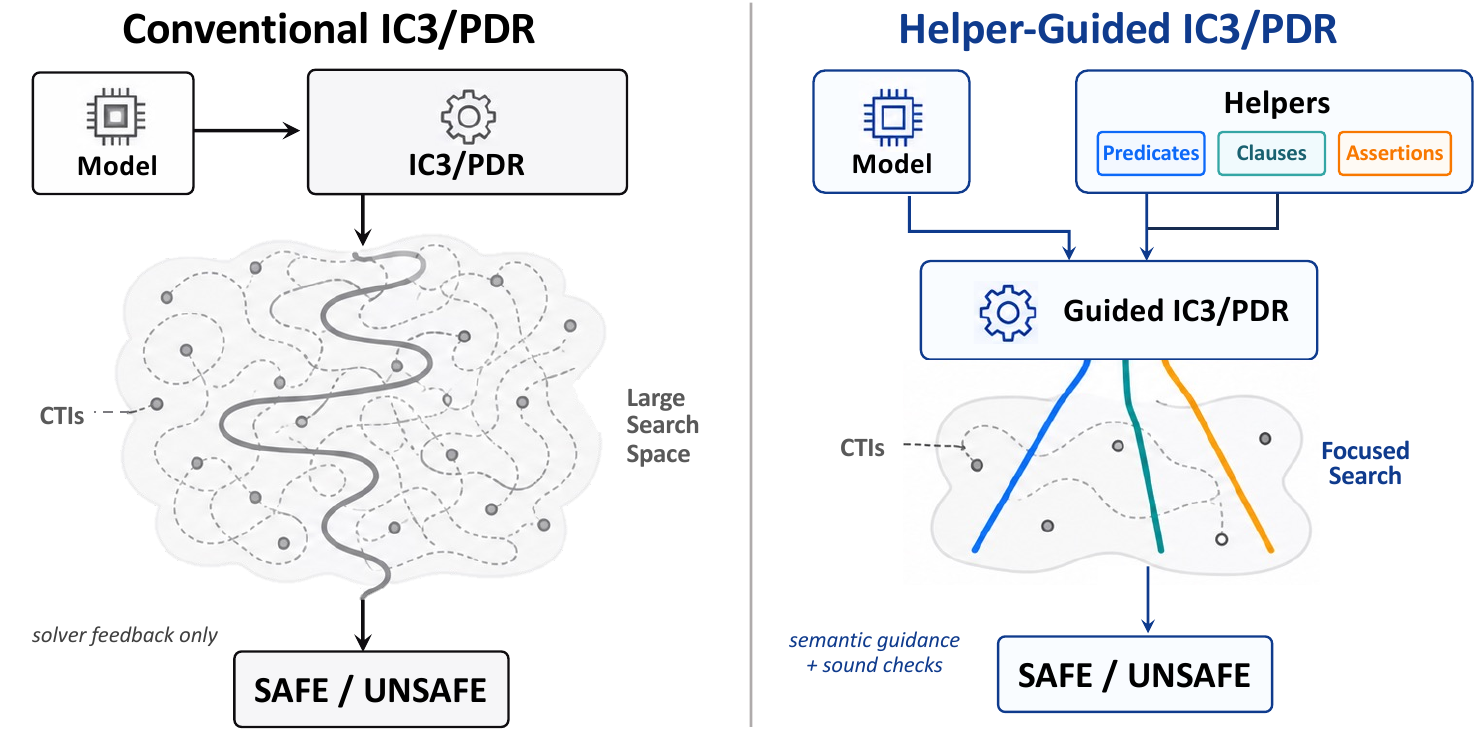}
    \caption{Motivation for helper-guided PDR. Conventional PDR learns blocking clauses mainly from solver feedback during the proof search. \texttt{LLM4PDR} adds verifier-checked semantic helpers at the predicate, clause, and assertion levels, guiding the same proof search toward word-level relations while preserving soundness through model-checker-side checks.}
    \label{fig:intro_comparison}
\end{figure}

Word-level PDR addresses this limitation by reasoning directly over word-level theories, such as bit-vectors and arrays. A simple arithmetic relationship between two variables, which may require a long chain of Boolean clauses after bit-blasting, can be represented as a single word-level formula. This compact representation gives word-level PDR a natural advantage on datapath-heavy designs. At the same time, the verifier may operate on an intermediate transition-system representation such as BTOR2~\cite{btorspec}, where source-level cues such as module boundaries, naming conventions, structural intent, and comments are reduced or absent. A relation that is otherwise apparent in the RTL can become hard for the proof engine to discover.

This information loss creates a concrete obstacle for PDR. During the proof search, PDR learns blocking clauses from solver feedback and then tries to generalize them so that each learned clause rules out a number of unreachable states at once. However, the feedback available to the engine typically expose only flat bit-vector terms or local signal assignments. For example, the proof may need the relation that two counters advance with a fixed offset, while the solver only exposes pairs of allowed values and it is difficult to discover a higher-level relation from such low-level value pairs. As a result, the engine may learn many weak local clauses instead of a compact word-level lemma.

In this paper, we study the following research question: can external semantic insight of the hardware design guide the PDR engine toward useful word-level facts without compromising the soundness of the algorithm? Today's Large Language Models (LLMs) are well suited for generating such semantic hints because they have been trained on large corpora of source-level hardware artifacts, including digital logic circuits modeled in hardware description languages, generator-produced hardware code, and formal specifications. They can recognize recurring design idioms (such as counters, pipeline stages, and handshake protocols), and propose candidate relations that those idioms suggest. However, these suggestions must not be trusted as proofs. To preserve soundness, the LLM can only act as a hint generator; the model checker remains responsible for accepting, rejecting, or refining every generated artifact.\looseness=-1

This paper introduces \textbf{\texttt{LLM4PDR}}, a framework that uses LLM-generated semantic hints to guide word-level PDR while leaving all proof decisions inside the verifier. \Cref{fig:intro_comparison} illustrates this shift: conventional PDR relies on solver feedback to discover blocking clauses, whereas \texttt{LLM4PDR} augments the same proof search with verifier-checked semantic helpers. The key observation is that semantic guidance can be useful at three verifier-side interfaces in the PDR workflow.
First, before a bad-state proof obligation is generalized, a \emph{helper predicate} can add a word-level atom to the vocabulary from which PDR builds blocking lemmas. Second, before the main blocking loop starts, a \emph{helper clause} can be checked as a candidate frame lemma. Third, during property checking, a \emph{helper assertion} can temporarily strengthen the target property and then be removed if counterexamples show that it is invalid. These three roles correspond to \emph{predicate}, \emph{clause}, and \emph{assertion}\footnote{Informally speaking, a \emph{predicate} is a Boolean condition over state variables, such as $x<y$; a \emph{clause} is a disjunction of such conditions or their negations, often learned by PDR as a lemma; an \emph{assertion} is a safety property, or an auxiliary fact conjoined with it, that the verifier attempts to prove.} guidance, instantiated as LLM-guided predicate generation, clause generation, and assertion generation, respectively. In all cases, the LLM output is untrusted: generated artifacts are parsed, type-checked, validated, and either used or discarded by the model checker. 

Our contributions are summarized as follows.
\begin{itemize}
    \item \textbf{Semantic-guided word-level PDR framework.} We propose \texttt{LLM4PDR}, a framework that utilizes LLM-generated semantic hints to guide word-level PDR without trusting the LLM as a proof oracle.
    \item \textbf{Three verifier-side guidance mechanisms.} We identify three natural injection points in PDR and instantiate them as predicate-level guidance for inductive generalization, clause-level guidance for frame strengthening, and assertion-level guidance for property strengthening.
    \item \textbf{Unified helper interface and validation pipeline.} To support these mechanisms uniformly, we design a restricted SMT-LIB2~\cite{smt2} helper interface and a verifier-side pipeline that parses, type-checks, validates, and refines LLM-generated artifacts before they can affect the proof search.
    \item \textbf{Implementation and evaluation.} We implement \texttt{LLM4PDR} in Pono~\cite{pono} and evaluate it on arithmetic, HLS-generated, open-source RTL, and HWMCC benchmarks. The results show complementary benefits across different structural classes, including a higher number of solved arithmetic cases and large speedups or timeout avoidance on several hard word-level instances.
\end{itemize}

The remainder of this paper is organized as follows. Section~\ref{sec:preliminary} reviews the IC3/PDR algorithm and problem formulation. Section~\ref{sec:motivation} motivates the need for LLM-guided reasoning with a concrete example. Section~\ref{sec:method} presents the three LLM-guided mechanisms and their integration into the PDR engine. Section~\ref{sec:experiment} reports experimental results across four benchmark suites, including ablation studies that isolate the contribution of each mechanism. Finally, Section~\ref{sec:conclusion} concludes the paper.

\section{Preliminaries and Problem Formulation}
\label{sec:preliminary}

\subsection{Symbolic Transition Systems}
\label{sec:transition_systems}
A finite-state transition system $S$ is defined by a tuple $(X, Y, I, T)$, where $X$ is a set of state variables, $Y$ is a set of primary input variables, $I(X)$ is a formula representing the initial states, and $T(X, Y, X')$ is the transition relation, mapping current states and inputs to next states. A state can be regarded as an assignment to all variables in $X$, often represented as a cube (a conjunction of literals). A safety property $P(X)$ asserts that only $P$-states (those whose assignments make $P$ evaluate to true) are reachable. The model checking problem asks whether all reachable states of $S$ satisfy $P$. 
For a formula $\varphi$ , we use the form $\varphi'$ to denote the case when each of its variable $x \in X$ is replaced by the primed version $x' \in X'$.

\subsection{The PDR Algorithm}
\label{sec:ic3_background}
PDR (a.k.a. IC3)~\cite{bradley2011ic3,emr-pdr-2011} is a SAT-based algorithm that proves safety properties by incrementally constructing an inductive invariant. The algorithm maintains a sequence of over-approximating frames $F_0, F_1, \ldots, F_k$. Each frame $F_i$ is a formula (a set of clauses that are conjuncted) whose models over-approximate the set of states reachable in at most $i$ steps. The frame sequence satisfies the following invariants:
\begin{itemize}
    \item $F_0 = I$. \hfill (initiation)
    \item For all $0 \le i < k$:
    \begin{enumerate}
        \item $F_i \models F_{i+1}$. \hfill (monotonicity)
        \item $F_i \land T \models F'_{i+1}$. \hfill (consecution)
        \item $F_i \models P$. \hfill (safety)
    \end{enumerate}
\end{itemize}

A counterexample to induction (CTI) is a state that is admitted by the current over-approximation but violates the target property. 
When a CTI is checked to be unreachable, a lemma will be conjuncted to the frame where the CTI was found.
The structure of the frame gives the proof object a simple hierarchy. A \emph{predicate} is a Boolean atom over state variables, such as $y=x+2$. A literal is a predicate or its negation, and a clause is a disjunction of literals. Since each frame is in the conjunctive normal form (CNF), a frame is, therefore, a conjunction of clauses. A proof is built gradually by adding clauses to frames; each clause may contain one or more predicates, with a unit clause being the special case that contains a single predicate.

The algorithm operates by alternating between a \emph{blocking} phase and a \emph{propagation} phase. Algorithm~\ref{alg:ic3_main} outlines the high-level procedure. In the blocking phase, PDR searches for a CTI, namely a predecessor $s$ of $\neg P$ in $F_k$, i.e., $F_k \land T \land \neg P'$ is satisfiable (Line 5). If such a state exists, the algorithm attempts to block it by recursively proving it unreachable from previous frames (Line 6). If the backward search reaches an initial state (Line 7), the property is violated. Otherwise, the propagation phase pushes clauses forward to subsequent frames (Line 8-12). If two consecutive frames become identical ($F_i = F_{i+1}$), a fixpoint is reached, and $F_i$ constitutes an inductive invariant proving the property (Line 13).

\begin{algorithm}[t]
\caption{High-level IC3/PDR Algorithm}
\label{alg:ic3_main}
\begin{algorithmic}[1]
\Procedure{PDR}{$I, T, P$}
    \If{$\text{IsSat}(I \land \neg P)$} \Return \textbf{false} \EndIf
    \State $F_0 \gets I$; $k \gets 1$; $F_k \gets P$
    \While{\textbf{true}}
        \While{$\text{IsSat}(F_k \land T \land \neg P')$} \Comment{Blocking Phase}
            \State $s \gets \text{GetState}(F_k \land T \land \neg P')$
            \If{\textbf{not} \Call{Block}{$s, k$}} \Return \textbf{false} \EndIf
        \EndWhile
        \State $k \gets k + 1$; $F_k \gets P$ \Comment{Propagation Phase}
        \For{$i \gets 1$ \textbf{to} $k-1$}
            \For{\textbf{each} clause $c \in F_i$}
                \If{\textbf{not} $\text{IsSat}(F_i \land c \land T \land \neg c')$}
                    \State $F_{i+1} \gets F_{i+1} \cup \{c\}$
                \EndIf
            \EndFor
            \If{$F_i = F_{i+1}$} \Return \textbf{true} \EndIf
        \EndFor
    \EndWhile
\EndProcedure
\end{algorithmic}

\end{algorithm}

\subsection{Proof Obligation Handling and Inductive Generalization}
\label{sec:inductive_generalization}
When PDR identifies a CTI $s$ at frame $i$, it generates a proof obligation $(s, i)$ and attempts to block it. Algorithm~\ref{alg:ic3_block_mic} illustrates this recursive blocking process.
The \textsc{Block} procedure checks if $s$ has a predecessor in $F_{i-1}$. If a predecessor $p$ exists, it recursively blocks $p$ at frame $i-1$. If no predecessor exists, $\neg s$ is inductive relative to $F_{i-1}$.

At this point, PDR performs \emph{inductive generalization} to strengthen $\neg s$ into a smaller clause $c \subseteq \neg s$ that remains inductive relative to $F_{i-1}$. This step is critical: a stronger clause (containing fewer literals) blocks a larger set of unreachable states, accelerating convergence. The minimal inductive clause (MIC) generalization~\cite{bradley2007checking} procedure iteratively attempts to drop literals from $\neg s$. For each candidate subclause $g$, the algorithm calls the \textsc{DOWN} subprocedure to verify and further minimize. 

\begin{algorithm}[t]
\caption{Recursive Blocking and Minimal Inductive Clause (MIC) Generalization}
\label{alg:ic3_block_mic}
\begin{algorithmic}[1]
\Procedure{Block}{$s, i$}
    \If{$i = 0$} \Return \textbf{false} \EndIf
    \While{$\text{SAT}(F_{i-1} \land \neg s \land T \land s')$}
        \State $p \gets \text{GetPredecessor}()$ \Comment{Extract predecessor state}
        \If{\textbf{not} \Call{Block}{$p, i-1$}} \Return \textbf{false} \EndIf
    \EndWhile
    \State $c \gets \Call{MIC}{\neg s, i}$ \Comment{Inductive Generalization}
    \For{$j \gets 1$ \textbf{to} $i$}
        \State $F_j \gets F_j \cup \{c\}$
    \EndFor
    \State \Return \textbf{true}
\EndProcedure

\Procedure{MIC}{$c, i$}
    \State $req \gets \emptyset$ \Comment{Literals that cannot be dropped}
    \For{\textbf{each} literal $l \in c$}
        \State $g \gets c \setminus \{l\}$
        \If{\Call{Down}{$g, i, req$}}
            \State $c \gets g$
        \Else
            \State $req \gets req \cup \{l\}$
        \EndIf
    \EndFor
    \State \Return $c$
\EndProcedure

\Procedure{Down}{$g, i, req$}
    \While{\textbf{true}}
        \If{$\text{SAT}(I \land \neg g)$} \Return \textbf{false} \EndIf
        \If{\textbf{not} $\text{SAT}(F_{i-1} \land T \land g \land \neg g')$} \Comment{Check consecution}
            \State $core \gets \text{UnsatCore}()$ \label{line:unsat_core}
            \State $g_{core} \gets \{ \text{literal } m \mid \neg m' \in core \}$
            \While{$\text{SAT}(I \land \neg g_{core})$} \Comment{Ensure initiation}
                \State pick $l \in g \setminus g_{core}$; $g_{core} \gets g_{core} \cup \{l\}$
            \EndWhile
            \State $g \gets g_{core}$ \Comment{Drop literals not in UNSAT core}
            \State \Return \textbf{true}
        \Else
            \State $s \gets \text{GetPredecessor}()$ \Comment{Extract predecessor state}
            \If{$(g \setminus \neg s) \cap req \neq \emptyset$} \Return \textbf{false} \EndIf
            \State $g \gets g \cap \neg s$ \Comment{Join operation to cover predecessor}
        \EndIf
    \EndWhile
\EndProcedure
\end{algorithmic}
\end{algorithm}

The \textsc{Block} procedure's \textbf{while} loop eventually terminates because each successful recursive call to \textsc{Block} adds a new clause to $F_{i-1}$, strengthening the frame and preventing the same predecessor $p$ from being extracted again. The \textsc{Down} procedure is the workhorse of MIC. It checks if $g$ satisfies consecution relative to $F_{i-1}$ using a SAT query. If the query is unsatisfiable (UNSAT), the SAT solver returns an \emph{UNSAT core}, namely a subset of the assumptions sufficient to cause the conflict. \textsc{Down} uses this core to simultaneously drop multiple unnecessary literals from $g$, significantly speeding up generalization. If the query is satisfiable (SAT), the solver has found a predecessor state $s$. To make the clause inductive, \textsc{Down} performs a \emph{join} operation ($g \gets g \cap \neg s$), which drops literals from $g$ to ensure the generalized cube covers the predecessor $s$.

The effectiveness of PDR heavily relies on the quality of the generalized clauses. As shown in Algorithm~\ref{alg:ic3_block_mic}, MIC's literal-dropping order is determined either by an arbitrary fixed order or by the SAT solver's UNSAT core extraction (Line~\ref{line:unsat_core}), with no semantic understanding of the literals' roles in the design. This reliance on SAT solver heuristics can result in generalizations that capture only local bit-level patterns, forcing the algorithm to explicitly enumerate and block many individual states, leading to deep backward searches and performance degradation~\cite{hassanBetterGeneralizationIC32013b, griggio2015comparing}. The same issue also appears at the representation level: a compact word-level fact may require a large number of bit-level clauses, making it difficult for PDR to discover and exploit the underlying word-level semantics.


\subsection{Clause and Literal Explosion in PDR}
\label{sec:literal_explosion}
Bit-level PDR represents each frame as a CNF formula over Boolean literals. This representation is sound and solver-friendly, but its size can grow significantly when the proof depends on a word-level relation. A simple relation, such as $a=b+c$ or $x-y=2$, is a single atom at the word level, but after bit-blasting, the same relation is now over the bit slices of the word-level variables, and PDR  faces two kinds of blow-up. First, no single  clause captures such a relation, so PDR has to learn \emph{many} clauses, each ruling out only a small set of states; this is the \emph{clause explosion}. Second, each such clause can also contain \emph{many} literals, 
which is the \emph{literal explosion}. Both effects make the proof larger and slow down every SAT query in PDR. 


Existing engines address this bottleneck mainly in two ways. One possibility is to remain at the bit level, exemplified by the work from the   Luka and Vizel, who proposed the PdrER algorithm~\cite{luka2025pdrer} that adds auxiliary variables and definitions to shorten the proof. On one buffer-allocation benchmark, ordinary PDR needs more than $65{,}000$ invariant clauses, while the proof with auxiliary definitions is far smaller.  However, PdrER generates these definitions according to fixed patterns rather than deriving them from the design semantics. 
The second approach reasons directly at the word level, as exemplified by AVR~\cite{avr}. Instead of bit-blasting the design, AVR derives additional terms and predicates from the transition system, symbolic post-images, interpolants, and predefined deduction rules. Although this approach can potentially avoid bit-level blow-up, its terms and predicates are still generated by hand-crafted rules and built-in heuristics. Consequently, both approaches may fail to discover relations unique to a particular design: an engine may explore many irrelevant terms before finding the crucial relation, or may never find it at all. \Cref{sec:motivation}  illustrates such a case.


On the other hand, \texttt{LLM4PDR} takes a different approach. Instead of generating candidate facts from fixed rules, it uses an external LLM to read the static design (the RTL when it is available, and otherwise the BTOR2 transition system) and to propose helper predicates, clauses, or assertions in a strict SMT-LIB2 format. These suggestions are not trusted as invariants or assumptions. A helper predicate is used only as a candidate atom during generalization; a helper clause must pass frame-consistency checks before it is added to a frame; and a helper assertion is kept only if it survives counterexample-guided refinement. In this way, a relation that would otherwise need many bit-level clauses can enter the proof as one compact suggestion, while the verifier still makes every decision that affects soundness. 

\section{Motivation}
\label{sec:motivation}

\begin{figure}[!tb]
    \centering
    \begin{minipage}{0.82\columnwidth}
    \begin{Verbatim}[fontsize=\scriptsize, frame=single, framesep=1.5mm, commandchars=\\\{\}]
\vkw{module} top(\vdecl{input} clk, \vdecl{input} rst, \vdecl{input} lden,
           \vdecl{input} ps, \vdecl{input} [15:0] ia,
           \vdecl{input} [15:0] ib);
    \vdecl{reg} [15:0] ra, rb, rai, rbi;
    \vkw{always} @(\vkw{posedge} clk) \vkw{begin}
    \vkw{if} (rst) \vkw{begin}
        ra <= \vlit{0}; rb <= \vlit{0}; 
        rai <= \vlit{0}; rbi <= \vlit{0};
    \vkw{end} \vkw{else} \vkw{begin}
        \vkw{if} (lden) \vkw{begin}
        ra <= ia; rb <= ib; 
        rai <= ia; rbi <= ib;
        \vkw{end} \vkw{else} \vkw{begin}
        \vkw{if} (~ps) \vkw{begin}
            rai <= rai + \vlit{1}; rbi <= rbi + \vlit{1};
    \vkw{end} \vkw{end} \vkw{end} \vkw{end}
    \vkw{assert property} \vlit{(!(ra == 16'h1357 &&}
    \vlit{  rai == 16'h9356 && rb == 16'h2468 }
     \vlit{ && rbi == 16'h2467));}
    \vlit{// rai-ra = 16'h7fff, rbi-rb = 16'hffff}
\vkw{endmodule}
    \end{Verbatim}
    \end{minipage}
    \captionof{figure}{An example of a hard-to-prove property for existing SAT-based model checking algorithms.}
    \label{fig:motivating example}
\end{figure}

\begin{table}[b]
  \centering
  \caption{
  Results of different model checkers on the motivating example
  for word widths from 16 to 20 bits.
  }
  \label{case abc}

  \begin{tabular}{lc}
  \toprule
  Model checker & Result on 16--20-bit instances \\
  \midrule
  Berkeley-ABC~\cite{ABC}
      & \textcolor{red}{TO} on all widths \\
  rIC3~\cite{ric3}
      & \textcolor{red}{TO} on all widths \\
  AVR~\cite{avr}\textsuperscript{a}
      & \textcolor{red}{TO} on all widths \\
  PdrER~\cite{luka2025pdrer}
      & \textcolor{red}{TO} on all widths \\
  \bottomrule
  \end{tabular}

  \par\vspace{0.3em}
  \begin{minipage}{0.9\linewidth}
  \footnotesize
  \textbf{TO}: runtime exceeded the 3600-second limit for every
  tested width.
  \textsuperscript{a} The better result of the SA and
  SA+UF configurations is reported.
  \end{minipage}
\end{table}

\begin{figure*}[!t]
    \centering
    \includegraphics[width=1\textwidth]{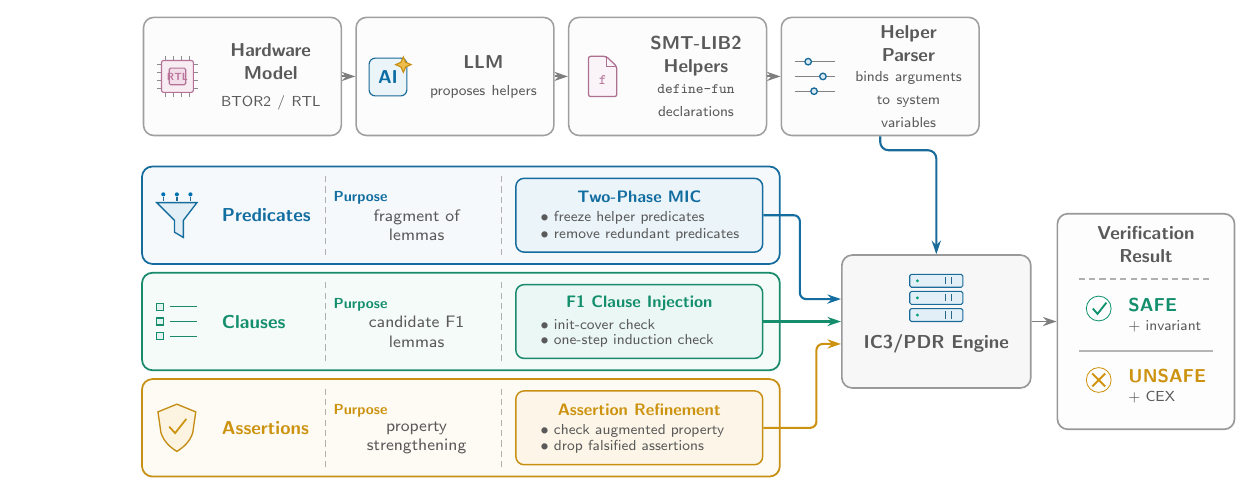}
    \caption{Overview of \texttt{LLM4PDR}. A hardware model is analyzed by an LLM or external tool to produce restricted SMT-LIB2 helper declarations. After parsing and binding helper terms to system variables, the verifier consumes predicates through two-phase MIC, clauses through $F_1$ injection, and assertions through property-strengthening refinement before returning a safe proof with an invariant or an unsafe counterexample.}
    \label{fig:framework_overview}
\end{figure*}

This section demonstrates a concrete example that remains challenging for the existing bit-level and word-level PDR engines.

One recurring challenge in hardware model checking is the \textit{discovery} of effective, high-level inductive invariants required to prove a property. PDR provides a robust framework for formal proof, yet prior work has shown that its performance depends heavily on the quality of inductive generalization and on the auxiliary predicates or lemmas available to the engine~\cite{bradley2007checking,hassanBetterGeneralizationIC32013b,griggio2015comparing,avr,luka2025pdrer}. Existing approaches derive such information through pre-coded rules, fixed templates, solver-feedback heuristics, or artifacts such as Craig interpolants~\cite{goel2019model,zhang2021syntax,mcmillan2003interpolation}.
But this strategy exposes a practical bottleneck: when the necessary arithmetic relationship is implicit in the update logic and does not match the engine's built-in generation patterns, the search for an inductive invariant can stagnate. 
Consider the property checking problem in \Cref{fig:motivating example} as an example.
The design contains four 16-bit state variables ($ra$, $rai$, $rb$, $rbi$) and a safety property asserting that a particular state assignment is unreachable.
Observe that $rai$ and $rbi$ are both incremented under identical guard conditions; consequently, the invariant $rai - ra = rbi - rb$ holds across all reachable states. A human designer recognizes this word-level relationship almost immediately. Yet at the bit level the subtraction $rai - ra$ alone expands into a chain of full-adder carry propagation across all 16 bits, producing an exponential blowup in the Boolean representation. A SAT-based engine must rediscover this arithmetic structure and express it using the propositional logic, whose formula size scales poorly  with word width. 
As \cref{case abc} shows, Berkeley-ABC~\cite{ABC}, rIC3~\cite{ric3}, AVR~\cite{avr}, and PdrER~\cite{luka2025pdrer} all fail to solve the instance within the one-hour timeout for every width from 16 to 20 bits.

Once a relation such as $rai - ra = rbi - rb$ is proposed, the verifier can check it using ordinary initiation and consecution queries. The hard part is that existing engines often do not synthesize this relation from the low-level encoding by themselves. This disparity calls for formal reasoning at a higher level. \looseness=-1

This limitation motivates augmenting rule-based lemma generation with externally proposed semantic hints. In this paper, we use large language models (LLMs) as semantic hint generators: given a static hardware description, the LLM can suggest proof-relevant relationships among state variables and emit candidate helper artifacts in a restricted SMT-LIB2 format that the verification engine can consume. This also differs from our prior work DreamMiner~\cite{dreamminer}, which learns property augmentations from simulation traces and dynamic execution data, while \texttt{LLM4PDR} uses just the static design information. 

\texttt{LLM4PDR} differs from fixed-rule word-level engines as the LLM is not limited to a pre-coded set of term-construction templates. It can use naming, structural context, and recurring RTL idioms to propose predicates, clauses, or assertions that are apparent to a designer but difficult for a heuristic term generator to synthesize. 
Importantly, this approach positions the LLM strictly as a semantic \textit{guide} rather than a trusted proof oracle. Because the verifier independently checks every generated candidate using its mathematical proof obligations, our framework can accelerate the proof search while preserving the soundness of the verification process.


\begin{figure}[!tb]
\centering
\begin{minipage}{0.98\columnwidth}
\begin{mdframed}[linewidth=0.6pt,linecolor=blue!55!black,backgroundcolor=blue!3,roundcorner=2pt,innertopmargin=4pt,innerbottommargin=4pt,innerleftmargin=5pt,innerrightmargin=5pt]
\begin{Verbatim}[fontsize=\scriptsize, commandchars=\\\{\}]
\vcmt{; helper predicate: extra MIC predicates}
(\vkw{define-fun} \vdecl{|predicate.k|} ((v1 S1) ... (vn Sn)) \vtype{Bool}
  \vlit{<SMT-LIB2 Boolean term>})

\vcmt{; helper clause: candidate frame lemma}
(\vkw{define-fun} \vdecl{|clause.k|} ((v1 S1) ... (vn Sn)) \vtype{Bool}
  \vlit{<SMT-LIB2 Boolean term>})

\vcmt{; helper assertion: tentative property strengthening}
(\vkw{define-fun} \vdecl{|assertion.k|} ((v1 S1) ... (vn Sn)) \vtype{Bool}
  \vlit{<SMT-LIB2 Boolean term>})
\end{Verbatim}
\end{mdframed}
\end{minipage}
\caption{Restricted SMT-LIB2 helper format. Each entry is an SMT-LIB2 \texttt{define-fun}; the reserved prefix determines how the PDR engine may consume the term.}
\label{fig:helper-dsl}
\end{figure}

\begin{figure*}[!t]
\centering
\includegraphics[width=\textwidth]{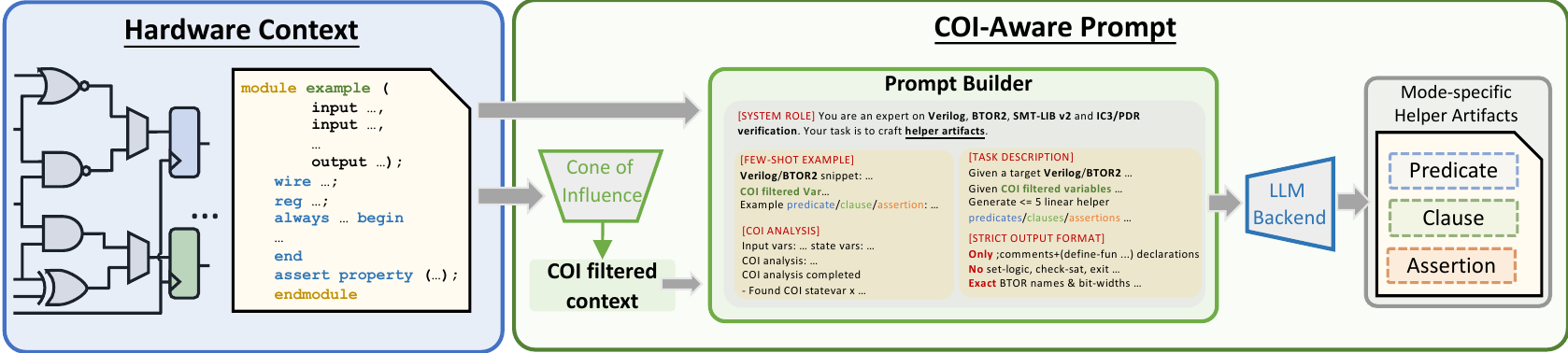}
\caption{COI-aware prompt construction and helper synthesis in \texttt{LLM4PDR}.
Given a verification problem, \texttt{LLM4PDR} uses RTL when available and otherwise the BTOR2 transition-system input.
A cone-of-influence (COI) analysis extracts property-relevant variables and transition fragments, which are inserted into a fixed prompt template together with the system role, a few-shot helper example, a mode-specific task description, and strict SMT-LIB2 output constraints.
The LLM backend then emits mode-specific helper artifacts, namely \texttt{predicate.*}, \texttt{clause.*}, or \texttt{assertion.*} declarations, which are later parsed, type-checked, and validated by the verifier before use.}
\label{fig-prompt-example}
\end{figure*}

\section{Methodology}
\label{sec:method}

Our framework, \texttt{LLM4PDR}, enhances Property Directed Reachability by integrating LLM guidance at three stages of the verification workflow. As shown in \Cref{fig:framework_overview}, an LLM extracts high-level insights from the RTL description and emits them as restricted SMT-LIB2~\cite{smt2} helper declarations (when RTL is not available, BTOR2 transition-system input is used instead). These insights become formal artifacts (predicates, clauses, and assertions), steering the PDR engine toward a more efficient proof search. 


\subsection{Restricted SMT-LIB2 Helper Interface}
\label{ssec:helper-dsl}

Before discussing individual mechanisms, we first describe the interface, through which the helper information enters the model checker. \texttt{LLM4PDR} uses a helper interface based on a restricted SMT-LIB2 file, rather than a separate domain-specific language. The LLM is expected to generate typed SMT-LIB2 definitions, where the reserved name prefixes tell the verifier the desired use cases of the hints. 
The interface bares no guarantee that the hints are useful and it does not modify any part of the transition systems. It is rather a side input that carries structured hints into the PDR engine.
%




\Cref{fig:helper-dsl} shows the overall structure of this SMT-LIB2 file. The SMT-LIB2 format is beneficial, because the  terms must already have precise bit-vector sorts and solver-level semantics under such format. The only extra convention is the reserved name prefixes. A \texttt{predicate.*} prefix indicates that the term follows should be placed in the helper-predicate pool $\mathcal{P}_{ext}$ and can be used as an additional predicate during MIC. A \texttt{clause.*} term is placed in the helper-clause pool $\mathcal{C}_{ext}$ and must respect the initiation, monotonicity and consecution requirements mentioned in Section~\ref{sec:ic3_background} when it is used to strengthen a frame. An \texttt{assertion.*} term is placed in the helper-assertion pool $\mathcal{A}_{ext}$ and is used as a tentative strengthening of the target property, with counterexamples used to remove helpers that turn out to be invalid. 


The distinction among helper predicates, helper clauses, and helper assertions follows the proof hierarchy from Section~\ref{sec:ic3_background}. A helper predicate is an atom that MIC may use when it constructs a blocking clause for one proof obligation; it serves as a candidate atom for generalization. 
A helper clause is a candidate frame lemma. Before insertion into $F_1$, it must pass such checks: the initial state predicate must imply the clause, and every one-step successor of an initial state must satisfy the primed clause. 
A helper assertion is a candidate strengthening of the target property and is filtered by counterexample-guided refinement. Thus the same word-level fact, such as $y=x+2$, can appear in three roles: as a predicate atom available to MIC, as a unit clause that can be checked for $F_1$ insertion, or as an assertion conjoined with the target property. 

The processing path follows three steps. First, the parser checks that the helper file is syntactically valid SMT-LIB2. Second, elaboration binds every formal argument in a helper term to a state variable in the transition system and checks that their bit-vector widths match. Third, the prefix dispatches the term to the corresponding validation path. This structured interface gives the LLM a concrete way to communicate word-level information, while keeping soundness inside the verifier: helper information can guide search, but the predicates, clauses, and augmented properties all remain subject to verifier-side checks. 

\subsection{Prompt Construction and Helper Synthesis}
\label{ssec:prompt-construction}

The helper interface above defines what the verifier can consume, while prompt construction determines how \texttt{LLM4PDR} asks the LLM to produce such helper artifacts. As shown in \Cref{fig-prompt-example}, \texttt{LLM4PDR} builds a COI-aware prompt from a static hardware context and sends the instantiated prompt to an LLM backend, which returns mode-specific helper artifacts. For each target property, \texttt{LLM4PDR} starts from the hardware representation available to the verifier, namely the relevant RTL fragment when RTL is available, and otherwise the BTOR2 transition-system input. The prompt also includes the target property and the state variables that may participate in proof-relevant helper declarations.

Before assembling the prompt, \texttt{LLM4PDR} applies cone-of-influence (COI) analysis with respect to the target property. COI identifies the state variables and transition fragments whose fan-in can affect the property. Variables and transition fragments outside this slice are omitted from the prompt, leaving the COI-filtered design context shown in \Cref{fig-prompt-example}. 
This filtering reduces the prompt size and serves as a property-directed relevance filter, restricting the model context to design elements that can structurally influence the target property.
It will not change the transition system, the target property, or any proof obligation checked by PDR. 
Empirically speaking, on representative HWMCC instances, this pruning removes between 52.35\% and 82.49\% of the state variables from the prompt, reducing both the prompt size and the space of irrelevant candidate helpers the model can propose.

The prompt builder then instantiates a fixed prompt template, where the system role asks the model to act as a hardware-verification assistant and to emit helper artifacts in the restricted SMT-LIB2 format. The few-shot example provides a small RTL or BTOR2 fragment together with correctly typed helper declarations, so the model sees the expected types, reserved name prefixes, and the overall output structure before it generates helpers for the target instance. The COI analysis  supplies the property-directed design context, including the selected variables and transition fragments. The task description selects the guidance mode and asks the model to generate helper predicates, helper clauses, or helper assertions. Finally, the strict output-format instruction constrains the requirements for  the SMT-LIB2 structure.

In each run, the mode-specific task in \Cref{fig-prompt-example} asks for up to five declarations with one reserved prefix: \texttt{predicate.*}, \texttt{clause.*}, or \texttt{assertion.*}. 
The prefix dispatches each declaration to the corresponding verifier-side path introduced in Section~\ref{ssec:helper-dsl}, where it is parsed, type-checked, and validated before it is used in the PDR procedure. Across all experiments, we use the same prompt template and per-request cap, so the three settings \texttt{LLM2Predicate}, \texttt{LLM2Clause}, and \texttt{LLM2Assert} that will be detailed in the next three subsections differ only in how the verifier consumes the generated information.

\subsection{LLM-Guided Predicate Generation}

\begin{figure}[!tb]
\centering
\begin{minipage}{0.58\columnwidth}
\begin{Verbatim}[fontsize=\scriptsize, frame=single, framesep=1.5mm, commandchars=\\\{\}]
\vkw{module} xp2(clk);
   \vdecl{input} clk;
   \vdecl{reg} [5:0] x = \vlit{6'd0};
   \vdecl{reg} [5:0] y = \vlit{6'd2};

   \vdecl{wire}      stop;
   \vkw{assign} stop = (x == \vlit{6'd61});

   \vkw{always} @(\vkw{posedge} clk) \vkw{begin}
      \vkw{if} (!stop) \vkw{begin}
         x <= x + \vlit{1};
         y <= y + \vlit{1};
      \vkw{end}
   \vkw{end}

   \vkw{always} @* \vkw{begin}
      x_lt_y: \vkw{assert}(x < y);
   \vkw{end}

\vkw{endmodule} \vcmt{// xp2}
\end{Verbatim}
\end{minipage}
\caption{RTL code of the \texttt{xp2} counter. The useful proof helper is a word-level relation between $x$ and $y$, but after translation into the bit level, the proof obligations expose low-level bit-vector terms rather than the word-level relations.}
\label{fig:xp2-rtl}
\end{figure}

\label{ssec:llm-predicate}

To help explain how helper predicates are used in our method, we start with a small example as shown by \Cref{fig:xp2-rtl}. It is not hard to find that a word-level inductive invariant $y = x + 2$ holds for $x$ and $y$ within the reachable-state bound $x\le61$ and it should be useful to prove the target property in this example. 

However, the bit-level IC3/PDR typically discovers CTIs as bit-level cubes: a state is represented by conjunction of literals, where each literal refer to a single bit in a multi-bit variable (written as a bit extraction). 
For example, a proof-obligation cube from an actual \texttt{xp2} run states that all six bits of $y$ are one and bit~1 of $x$ is one:
\[
    B = \{y[j]=1 \mid 0\le j\le5\}\cup\{x[1]=1\}.
\]
Here $v[i]$ denotes a single-bit slice of the bit-vector $v$, or term $\mathsf{extract}(i,i,v)$ in the SMT syntax. Here, we use $s$ to refer to the corresponding state formula $s \overset{\text{def}}{=} (y=63) \wedge (x[1]=1)$, and each bit assignment (such as $y[5]=1$) is in fact a literal.

As previously stated in Section~\ref{sec:inductive_generalization}, inductive generation attempts to find a clause $g$ to block a proof obligation $s$ at a certain frame $F_{i}$ when $s$ is shown to be unreachable from the prior frame $F_{i-1}$. To maintain the requirements of the frame sequence, the clause $g$ needs to satisfy the following requirements:
\begin{align}
                        I & \Rightarrow g \label{eq:g_initiation} \\
g \wedge F_{i-1} \wedge T & \Rightarrow g' \\
                        g & \Rightarrow \neg s \label{eq:g_blocks_neg_s}
\end{align}
The bit-level IC3/PDR implementation typically invokes the aforementioned MIC procedure (\Cref{alg:ic3_block_mic}) to remove the literals in $B$ to form the clause $g$.

However, those literals in CTI describe only one corner of the state space, and do not directly model the high-level arithmetic relation such as $y = x + 2$ that we need in the inductive invariant.
\begin{algorithm}[!tb]
\caption{Helper-Preserving Minimal Inductive Clause (MIC)}\label{alg:predicate_generation}
\begin{algorithmic}[1]
\Procedure{SynthesizePredicates}{$D$}
    \State $\mathcal{P}_{raw} \gets \text{QueryLLM}(D)$
    \State \Return $\{p \in \mathcal{P}_{raw} \mid \text{SAT}(p) \land \text{SAT}(\neg p)\}$ \Comment{filter out trivial predicates}
\EndProcedure
\Statex
\Procedure{TwoPhaseMIC}{$F_i, T, c, \mathcal{P}_{ext}$}
    \State $B \gets \text{BitLiterals}(c)$;\; $V_c \gets \text{Vars}(c)$
    \State $H \gets \emptyset$ \Comment{Helper literals determined by the CTI}
    \For{\textbf{each} $p(\vec{v}) \in \mathcal{P}_{ext}$ with $\vec{v} \subseteq V_c$} \label{line:match_start}
        \If{$c \models p$}
            \State $H \gets H \cup \{p\}$
        \ElsIf{$c \models \neg p$}
            \State $H \gets H \cup \{\neg p\}$
        \EndIf
    \EndFor \label{line:match_end}
    \State \textcolor{blue}{$B^\star \gets \Call{MinimizeBaseCube}{F_i, T, H, B}$} \Comment{Phase 1: $H$ is frozen} \label{line:phase1}
    \For{\textbf{each} helper literal $h \in H$} \Comment{Phase 2: Drop redundant helpers} \label{line:phase2start}
        \State $H_{tmp} \gets H \setminus \{h\}$
        \State $Q \gets H_{tmp} \cup B^\star$
        \State $\Phi \gets \text{Cube}(Q)$;\; $\Phi' \gets \text{Prime}(\Phi)$
        \If{$\text{SAT}(I \land \Phi)$}
            \State \textbf{continue} \Comment{Keep $h$: cube reaches initial states}
        \EndIf
        \If{\textcolor{blue}{$\text{SAT}(F_i \land \neg\Phi \land T \land \Phi')$}}
            \State \textbf{continue} \Comment{Keep $h$: no longer inductive}
        \EndIf
        \State \textcolor{blue}{$H \gets H_{tmp}$} \Comment{Drop redundant word-level helper}
    \EndFor \label{line:phase2end}
    \State \Return $\neg\text{Cube}(H \cup B^\star)$
\EndProcedure
\Statex
\Procedure{MinimizeBaseCube}{$F_i, T, H, B$}
    \State $\Phi \gets \text{Cube}(H \cup B)$
    \State \textcolor{blue}{$\Gamma \gets F_i \land \neg\Phi \land T \land \text{Cube}(H')$} \Comment{Hard constraints}
    \State \textcolor{blue}{$B^\star \gets \{b \in B \mid b' \in \Call{UnsatCore}{\Gamma; B'}\}$} \Comment{Only $B'$}
    \State \Return $\Call{RestoreInitiation}{I,H,B^\star,B}$
    \Statex \Comment{Restore literals as needed to exclude initial states}
\EndProcedure
\end{algorithmic}
\end{algorithm}

This is where the LLM is useful. We ask LLM to read the low-level design representation and generate a restricted SMT-LIB2 helper file with helper predicates (\textsc{SynthesizePredicates} in Algorithm~\ref{alg:predicate_generation} builds the candidate pool once for the design). For \texttt{xp2}, the generated helper file contains predicates such as:
\[
    p_1 : y = x + 2,\qquad
    p_2 : x < y,\qquad
    p_3 : x \leq 61.
\]
These predicates enrich the literal set available to the inductive generalization procedure. 
Because Formula~(\ref{eq:g_blocks_neg_s}) states that $g \Rightarrow \neg s$ and $g$ is a disjunction of literals, then every literal $l$ in $g$ should all follow the requirement that $s \Rightarrow \neg l$. Therefore, we may evaluate each available predicate $p$ under the CTI and use either $p$ or $\neg p$ as the candidate. If the outcome of the predicate is not fixed, then it means the candidate $p$ does not follow the above requirement and should be discarded for this inductive generalization attempt. This selection is described by Lines~\ref{line:match_start}--\ref{line:match_end} in Algorithm~\ref{alg:predicate_generation}.
In the running example, suppose the CTI state matches a helper-predicate subset $H_p$ from the generated file, so the working cube becomes:
\[
    C^+ = B \cup H_p.
\]

At first glance, one might simply give $C^+$ to ordinary MIC to select the literals. In practice, however, MIC may only select a few bit-level literals in $B$ while dropping all matched helper predicate candidates in $H_p$ entirely. This is because ordinary MIC sees every element of $C^+$ as a removable assumption, and the UNSAT core extraction in MIC has no direct control on which assumption to keep (it is totally dependent on the internal heuristics of the underlying SAT solver).

\begin{figure}[!tb]
\centering
\includegraphics[width=\columnwidth]{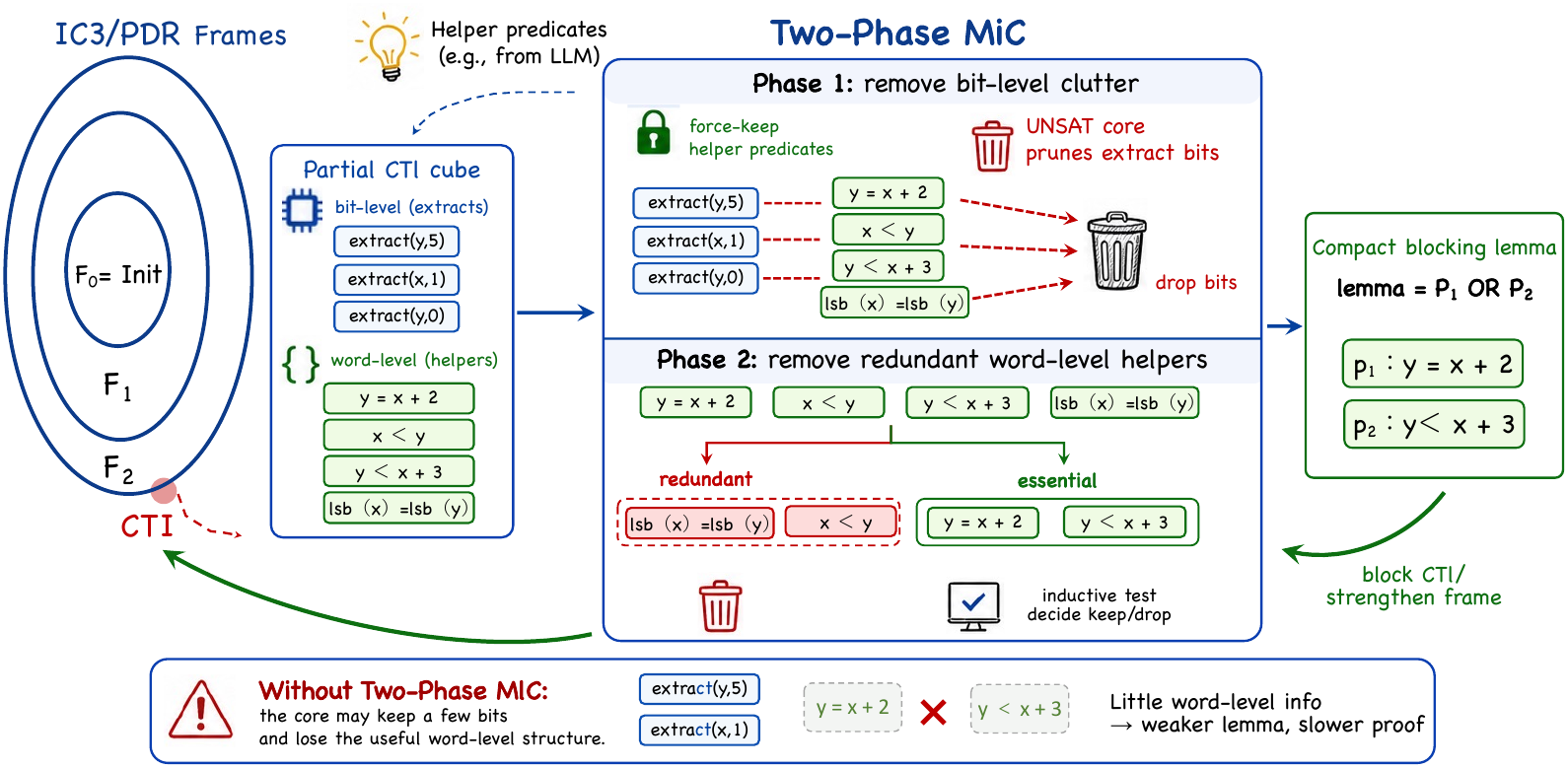}
\caption{Helper-predicate guided two-phase MIC. A CTI cube initially contains bit-level extract literals and matched word-level helper predicates. The first phase freezes helper predicates while pruning bit-level clutter; the second phase selectively removes redundant helper predicates through inductive checks, leaving a compact blocking lemma that preserves useful word-level structure.}
\label{fig:helper-predicate}
\end{figure}

Our solution is to make MIC work in two phases, as illustrated by \textsc{TwoPhaseMIC} in \Cref{fig:helper-predicate}. In the first phase (corresponding to Line~\ref{line:phase1} in \Cref{alg:predicate_generation}), the helper predicates in $H_p$ are frozen. The solver is allowed to remove only bit-level literals from $B$, while the matched helper literals remain hard constraints. Consequently, UNSAT-core minimization (\textsc{MinimizeBaseCube}) can shrink $B$ to a smaller set $B^\star$, but does not  discard any helper predicates. Operationally, the solver checks the usual initiation and consecution conditions while the candidate next-state helpers are asserted, so the core is allowed to explain only which bit-select literals are still needed.\looseness=-1


Then in the second phase (Lines~\ref{line:phase2start}--\ref{line:phase2end}), the helper predicates become removable.
The engine tries iteratively dropping helpers from $B^\star \cup H_p$. If removing a helper still excludes initial states and remains inductive relative to the current frame (both checks remain UNSAT), the helper predicate is deemed redundant and can be dropped. This lets MIC keep the generated word-level facts that are actually needed for the CTI, while discarding generated facts that are merely implied by the remaining cube. The final lemma is therefore compact and word-level, instead of being a handful of bit-selects.

Though LLM-generated predicates are used, the soundness of the IC3/PDR algorithm is preserved.  The LLM never gets to add a fact directly to a frame. All external predicates are checked to ensure the resulted clause $g$ follows the requirements stated in Formula~(\ref{eq:g_initiation})--(\ref{eq:g_blocks_neg_s}).


\subsection{LLM-Guided Clause Generation}
\label{ssec:llm-clause}

\begin{algorithm}[!tb]
\caption{Helper Clause Side-loading}\label{alg:clause_generation}
\begin{algorithmic}[1]
\Require Transition system $(\Init, T)$, PDR frames $F_0, F_1$, Helper clauses $\mathcal{C}_{cand}$
\Ensure $F_1$ strengthened with sanity-checked helper clauses
\Procedure{SideLoadClauses}{$\mathcal{C}_{cand}$}
\For{each clause $c \in \mathcal{C}_{cand}$}
    \If{$c$ is not Boolean} \textbf{continue}
    \EndIf
    \If{\textcolor{blue}{$\textsc{Sat}(\Init \land \neg c)$}} \textbf{continue} \Comment{violates $F_0 \Rightarrow F_1$}
    \EndIf
    \State $c' \leftarrow \Call{Prime}{c}$
    \If{\textcolor{blue}{$\textsc{Sat}(\Init \land T \land \neg c')$}} \textbf{continue} \Comment{violates $F_0 \land T \Rightarrow F_1'$}
    \EndIf
    \State \textcolor{blue}{$F_1 \leftarrow F_1 \land c$} \Comment{$c$ is a valid first-frame lemma}
\EndFor
\While{\texttt{true}} \Comment{Main PDR loop}
    \While{$F_k \land T \land \neg P'$ is SAT with witness $c$}
        \If{\textbf{not} \Call{Block}{$c$, $k$}}
            \State \Return \texttt{Unsafe}
        \EndIf
    \EndWhile
    \State $\cdots$ \Comment{Main PDR loop continues}
\EndWhile
\EndProcedure
\end{algorithmic}
\end{algorithm}

Helper clauses use the same LLM-proposed word-level relations at a different point in the PDR workflow. While helper predicates wait for a CTI and then give MIC an expanded set of candidate atoms, helper clauses are used earlier and more directly: before the main blocking loop starts, the verifier asks whether any generated formula is already strong enough to be placed in the first frame $F_1$. This is referred to as the clause side-loading mechanism in the prior work~\cite{deepic3}. It gives PDR a better initial frame sequence, without any change to either the transition system or the target property.

\begingroup
\postdisplaypenalty=10000
The distinction matters in the \texttt{xp2} example. Without side-loaded clauses, 
$F_1$ in the 
the initial frame sequence is set to the target property, namely $F_1 = P$, and the engine has to discover the arithmetic relation between $x$ and $y$ through CTIs. In practice, this produces many local bit-slice lemmas.
Clause generation asks the LLM to propose complete frame lemmas (clauses) up front. For \texttt{xp2}, the helper-clause file contains the desired relation
\[
    c_{\Delta}: y = x + 2,
\]
as well as other conjectures, such as
\[
    c_{\mathit{par}} : (x \mathbin{\&} 1) = (y \mathbin{\&} 1),
    \qquad
    c_i : (y < x+3) \lor (x[2] \neq 0).
\]
\endgroup
The formula $c_{\Delta}$ will be treated as a unit clause after being loaded, which exposes the whole proof idea: if $y=x+2$ is present in the frame, the property $x<y$ is no longer something PDR must reconstruct from unrelated bit patterns. The other clauses illustrate why clause generation is not merely predicate generation under another name. For example, $c_i$ is a disjunction-shaped frame lemma: it combines the arithmetic predicate $y<x+3$ with a bit-level condition on $x[i]$ and rules out a family of nearby unreachable states. Such clauses can be inserted and propagated as complete lemmas, even when they are not single word-level atoms.

Side-loading is only helpful when the injected clauses are good. A large set of irrelevant clauses can slow every SAT query and steer the search toward unproductive regions. Therefore \texttt{LLM4PDR} treats every formula in $\mathcal{C}_{ext}$ as a candidate, not as a trusted invariant. After parsing and type-checking, the engine performs the two sanity checks required for a clause to live in $F_1$:
\[
    \mathit{UNSAT}(\Init \land \neg c),
    \qquad
    \mathit{UNSAT}(\Init \land T \land \neg c').
\]
The first query checks that adding $c$ preserves $F_0 \Rightarrow F_1$: no initial state may violate the side-loaded clause. The second checks that $F_0 \land T \Rightarrow F_1'$: every one-step successor of an initial state must satisfy the primed clause. 

Algorithm~\ref{alg:clause_generation} summarizes such validation path. After a candidate passes both checks, the lemma object will be inserted into $F_1$. From that point on, the clause is handled by the ordinary PDR machinery. If it is relatively inductive at later frames, propagation will push it forward; if not, it stays behind and will not appear in the final inductive invariant. In the \texttt{xp2} run, the relation $y=x+2$ is side-loaded into $F_1$ and then propagates with the proof, while clauses that only satisfy the first-frame checks are naturally filtered by subsequent propagations.

The soundness boundary is the same as in helper-predicate generation. The LLM can propose a word-level lemma, but the verifier decides whether the lemma is legal at $F_1$, and the existing propagation checks decide whether it survives beyond $F_1$. Thus clause generation gives PDR a head start with the checks to ensure the frame sequence is sound.

\begin{algorithm}[!tb]
\caption{CEX-Guided Helper Assertion Refinement}\label{alg:assert_generation}
\begin{algorithmic}[1]
\Require Transition system $(\Init,T)$, target property $P$, helper assertions $\mathcal{A}_{cand}$
\Ensure \texttt{Safe} for $P$, or \texttt{Unsafe} with a real counterexample
\Procedure{CheckWithAssertions}{$P,\mathcal{A}_{cand}$}
    \State \textcolor{blue}{$V \gets \mathcal{A}_{cand}$} \Comment{active helpers not yet invalidated} \label{line:assert_active}
    \State \textcolor{blue}{$P^\star \gets P \land \bigwedge_{a \in V} a$} \label{line:assert_aug}
    \While{\texttt{true}}
        \State $c \gets \Call{GetBadCube}{P^\star}$
        \If{$c \neq \bot$}
            \If{\textbf{not} \Call{Block}{$c$}}
                \If{\textcolor{blue}{$\Call{Eval}{P,c}=\texttt{false}$}} \label{line:assert_check_original}
                    \State \Return \texttt{Unsafe}
                \EndIf
                \State \textcolor{blue}{$V \gets \{a \in V \mid \Call{Eval}{a,c}=\texttt{true}\}$} \Comment{remove falsified helpers} \label{line:assert_filter}
                \State \textcolor{blue}{$P^\star \gets P \land \bigwedge_{a \in V} a$} \label{line:assert_rebuild}
                \State \Call{RestartWith}{$P^\star$}
            \EndIf
        \Else
            \State \Call{NewFrame}{}
            \If{\Call{Propagate}{}}
                \State \Return \texttt{Safe}
            \EndIf
        \EndIf
    \EndWhile
\EndProcedure
\end{algorithmic}
\end{algorithm}

\subsection{LLM-Guided Assertion Generation}
\label{ssec:llm-assertion}

Helper assertions use LLM-proposed word-level conjectures at the property level. 
While helper predicates and helper clauses do not change the transition system or the target property,
the helper assertion instead attempts to strengthen the target property temporarily. The verifier asks PDR to prove
\[
    P^\star = P \land \bigwedge_{a \in V} a,
\]
where $V \subseteq \mathcal{A}_{ext}$ is the set of helper assertions that have not yet been invalidated. Since $P^\star \Rightarrow P$, proving the stronger property proves the original one, while giving PDR a more informative property frame to build from.
Algorithm~\ref{alg:assert_generation} summarizes this counterexample-guided refinement workflow. Lines~\ref{line:assert_active}--\ref{line:assert_aug} initialize the active helper set and construct the strengthened property.

\begin{figure}[!tb]
\centering
\includegraphics[width=\columnwidth]{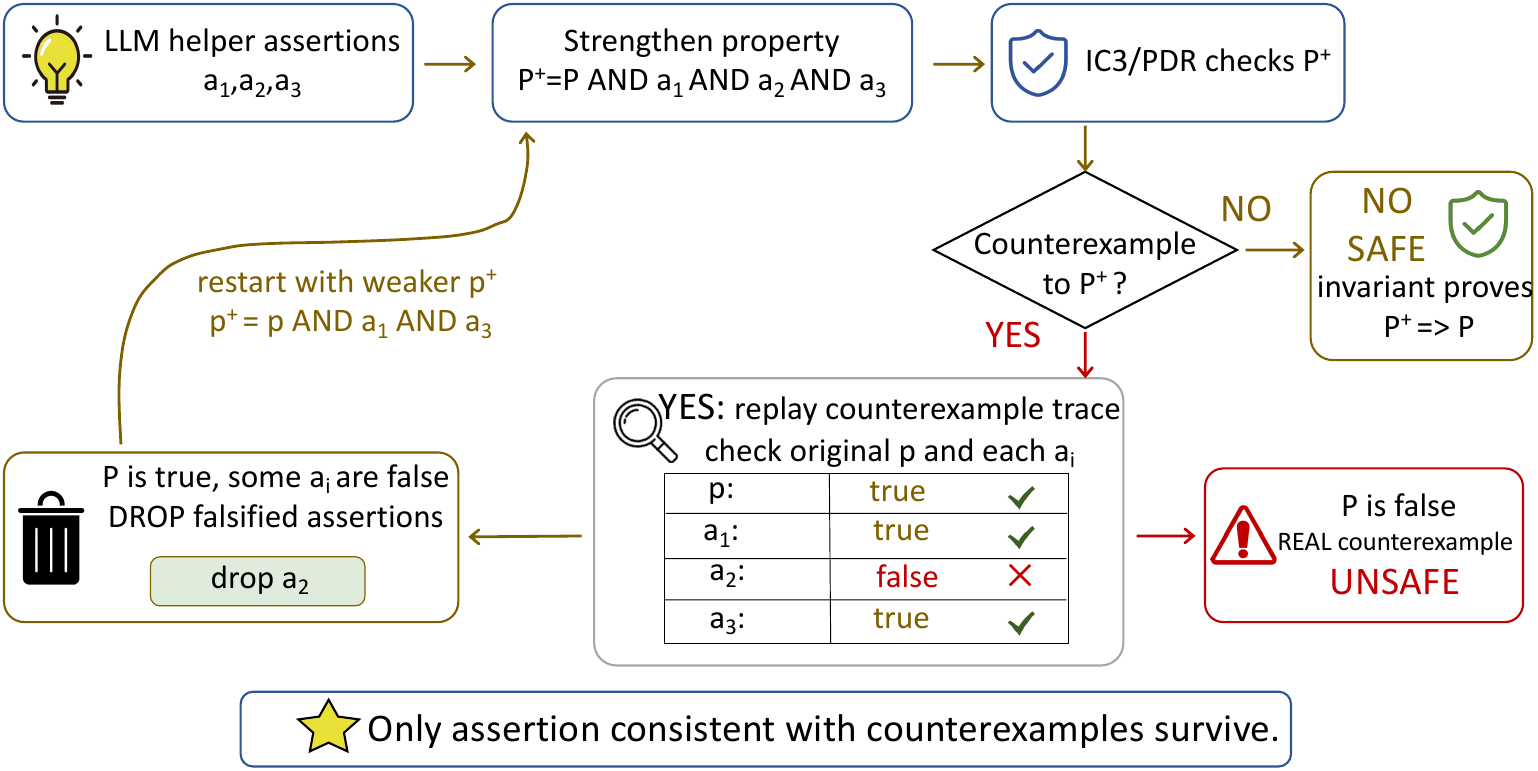}
\caption{Counterexample-guided refinement for helper assertions. The verifier checks the strengthened property $P^\star=P\land\bigwedge_{a\in V} a$. A counterexample to $P^\star$ is replayed against the original property: if it violates $P$, it is a real counterexample; otherwise, only the falsified helper assertions are removed and PDR restarts with a weaker strengthened property.}
\label{fig:helper-assert}
\end{figure}

The \texttt{xp2} example shows why this is useful. The target property is $P: x<y$. The LLM may propose the word-level relation
\[
    a_{\Delta}: (y-x)=2.
\]
This formula is the same design fact that can be used by the predicate and clause mechanisms, but it is consumed differently when it works as a helper assertion. With helper predicates, $(y-x)=2$ is only a candidate atom for MIC. With helper clauses, $(y-x)=2$ must pass frame checks before being inserted into $F_1$. 
With helper assertions, the verifier therefore uses the candidate to form the augmented property $P^\star=(x<y)\land ((y-x)=2)$, so the property frame already exposes the arithmetic relation that explains why $x<y$ holds.

This direct augmentation is powerful but unsafe if used blindly, because an LLM may propose a false assertion. We therefore follow the counterexample-guided refinement pattern in Algorithm~\ref{alg:assert_generation}. Whenever the strengthened property produces an unblockable bad cube $c$, the verifier first evaluates the original property $P$ on that cube (Line~\ref{line:assert_check_original}). If $P$ itself is false, the cube is a real counterexample and the design is unsafe. Otherwise, the counterexample only shows that the current helper set was too strong. The verifier removes exactly the helper assertions falsified by $c$ (Line~\ref{line:assert_filter}), rebuilds $P^\star$ from the remaining helpers (Line~\ref{line:assert_rebuild}), and restarts the ordinary PDR loop. \Cref{fig:helper-assert} pictorially illustrates how invalid helper assertions are filtered without compromising the original proof obligation.\looseness=-1

For \texttt{xp2}, the assertion $a_{\Delta}$ is true on all reachable states, so no refinement is needed. The strengthened property exposes the arithmetic reason behind $x<y$ immediately, and the proof can converge using the compact relation $y-x=2$ instead of rediscovering it through many bit-level CTIs. If the LLM had proposed a wrong relation, such as $y-x=3$, the first counterexample satisfying $x<y$ but violating that helper would simply delete the helper and continue with the original property.

The soundness boundary is again kept inside the verifier. Helper assertions are not added to the transition relation and are not trusted as assumptions. They are tentative property augmentations. If they help, they become part of a proof to the target property; if they are contradicted by a counterexample, they will be removed. In the worst case, all helper assertions are filtered out and Algorithm~\ref{alg:assert_generation} degenerates to ordinary PDR working on the original property $P$.

\section{Experimental Evaluation}\label{sec:experiment}
\begin{figure*}[t]
\centering
\includegraphics[width=\linewidth]{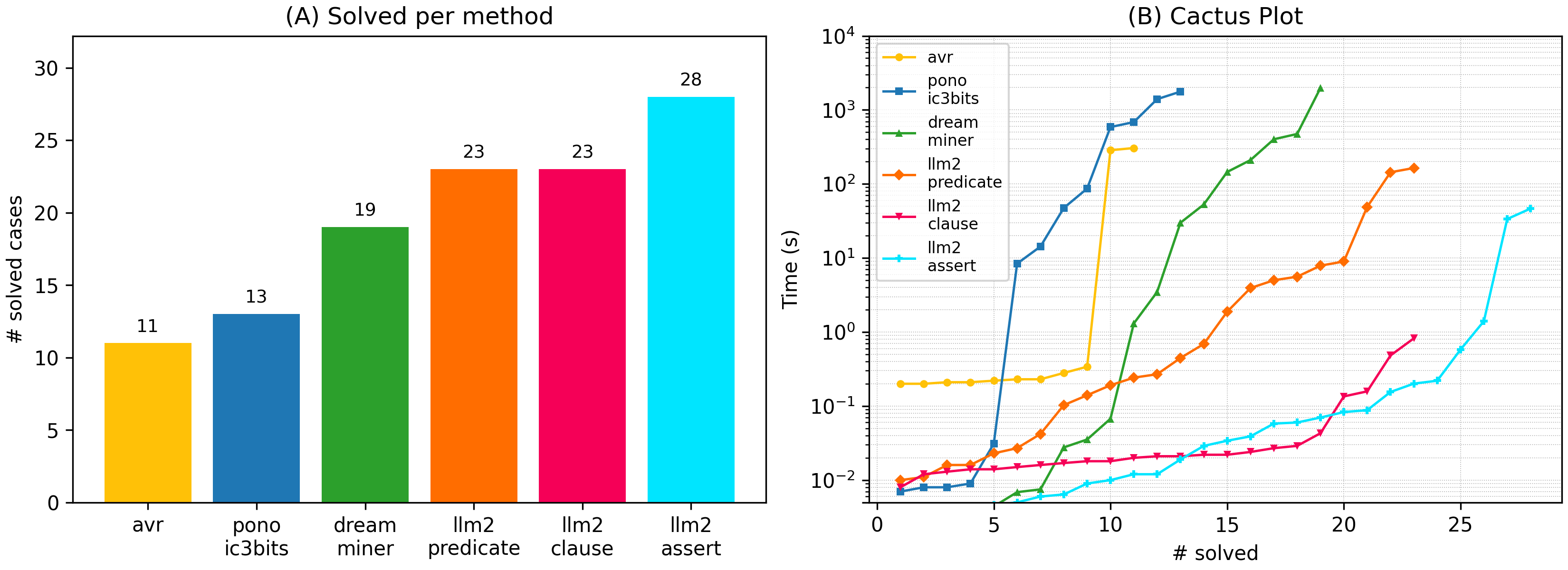}
\caption{Performance comparison on simple arithmetic benchmarks.}
\label{fig-simple-arith}
\end{figure*}
\subsection{Setup and Baselines}

We evaluate \texttt{LLM4PDR} on four benchmark suites that exercise different sources of difficulty in word-level hardware model checking. The arithmetic suite isolates datapath-heavy reasoning over additions, multiplications, shifts, and extracts. The HWMCC suite contributes larger competition benchmarks with mixed control and datapath structure. The HLS--SEC suite contains fixed-point Kalman-filter kernels generated by high-level synthesis, where the properties encode bit-exact equivalence between independently pipelined implementations. The open-source RTL suite contains formally annotated bus interconnects and DDR3 controllers, representing realistic control-rich SoC components.

For all LLM4PDR runs, we use OpenAI GPT o3~\cite{OpenAI2025o3}
as the LLM backend. We use a fixed prompt template across all benchmarks and do not perform benchmark-specific prompt tuning.
For each benchmark instance and each guidance mode, the LLM is asked to generate up to five helper artifacts in the restricted SMT-LIB2
helper format: five \texttt{predicate.*} declarations for \texttt{LLM2Predicate},
five \texttt{clause.*} declarations for \texttt{LLM2Clause}, or five \texttt{assertion.*} declarations for \texttt{LLM2Assert}. 

\subsection{Arithmetic Benchmarks}

Across the suites, we compare against Pono~\cite{pono} with the IC3bits engine (which uses the SMT-Switch backend to interface with the Bitwuzla SMT solver~\cite{niemetz2023bitwuzla}), AVR~\cite{avr} (the champion of the 2020's HWMCC), and DreamMiner~\cite{dreamminer} (a recent reinforcement-learning-based property mining and argumentation baseline). 
As  \texttt{LLM4PDR} is built upon a customized IC3 engine (IC3ng) that we implemented in Pono,
we also include an \emph{Ablation} configuration that runs Pono with IC3ng but disables all LLM-generated helper artifacts. This isolates the effect of the proposed guidance from the effect of changing the underlying PDR engine. We report three single-mechanism instantiations of \texttt{LLM4PDR}: \texttt{LLM2Predicate}, \texttt{LLM2Clause}, and \texttt{LLM2Assert}. They enable predicate generation, clause generation, and assertion generation, respectively, as described in Section~\ref{sec:method}. Runtimes are wall-clock times; entries marked \textcolor{red}{TO} hit the timeout limits described in the corresponding table notes. 

\begin{table*}[htp]
  \centering
  \scriptsize
  \begin{threeparttable}
  \renewcommand{\arraystretch}{1.12}
  \setlength{\tabcolsep}{1pt}
  \sisetup{
    table-format = 4.2,
    group-digits = false,
    detect-weight = true,
    detect-inline-weight = math
  }
  \caption{Benchmark sizes and runtime comparison on HWMCC19--24 benchmarks. Runtimes are in seconds.\tnote{a}}
\label{tab:hw-runtime}
  \begin{tabular*}{\linewidth}{@{\extracolsep{\fill}}
    >{\raggedright\arraybackslash}m{2.4cm}
    l
    c  
    S[table-format=2.2]  
    S  
    S  
    S  
    S  
    S  
    S  
  @{}}
    \toprule
    \textbf{Category} & \textbf{Benchmark} &
    \textbf{Size}\tnote{e} & \textbf{Pono\tnote{b}} & \textbf{AVR\tnote{c}} &
    \textbf{Ablation\tnote{d}} & \textbf{DreamMiner} & \shortstack{\texttt{LLM2}\\\texttt{Predicate}} & \shortstack{\texttt{LLM2}\\\texttt{Assert}} & \shortstack{\texttt{LLM2}\\\texttt{Clause}} \\
    \midrule
    \multirow{3}{=}{Accelerator} & marlann\_cp\_fail1\_p2 & \num{693}\,/\,\num{1441} & 26.64 & 2083.95 & 25.55 & 25.63 & 23.94 & \cellcolor{blue!10}\num{13.50} & 21.48 \\
    & marlann\_cp\_fail2\_p0 & \num{690}\,/\,\num{1441} & 23.23 & 315.02 & 6.92 & 9.07 & 6.46 & 8.39 & \cellcolor{blue!10}\num{6.20} \\
    & marlann\_cp\_pass\_p2 & \num{694}\,/\,\num{1441} & 26.58 & 2439.43 & 26.07 & 35.66 & 23.21 & \cellcolor{blue!10}\num{13.61} & 20.89 \\
    \midrule
    \multirow{3}{=}{Embedded CPU core} & picorv32\_p01 & \multirow{3}{*}{\num{1911}\,/\,\num{1878}} & 5.99 & \cellcolor{blue!10}\num{0.50} & 2.78 & 10.11 & 2.70 & 4.51 & 3.52 \\
    & picorv32\_p05 &  & 18.00 & \cellcolor{blue!10}\num{0.64} & 4.32 & 17.55 & 4.21 & 9.83 & 5.90 \\
    & picorv32\_p06 &  & 25.22 & \cellcolor{blue!10}\num{0.71} & 5.60 & 75.37 & 5.53 & 5.28 & 6.47 \\
    \midrule
    DSP / signal-estimation & intersymbol\_conv & \num{308}\,/\,\num{59} & {\color{red}TO} & {\color{red}TO} & 613.64 & {\color{red}TO} & \cellcolor{blue!10}\num{3.06} & 29.80 & 3445.22 \\
    \midrule
    \multirow{2}{=}{Peripheral / storage ctrl.} & qspiflash\_p164 & \num{3100}\,/\,\num{614} & {\color{red}TO} & 2553.76 & 1310.61 & {\color{red}TO} & 1220.00 & \cellcolor{blue!10}\num{859.80} & {\color{red}TO} \\
    & qspiflash\_p078 & \num{3098}\,/\,\num{613} & {\color{red}TO} & {\color{red}TO} & 299.92 & {\color{red}TO} & 77.27 & 100.50 & \cellcolor{blue!10}\num{53.89} \\
    \midrule
    Display / video interface & vgasim\_p89 & \num{2003}\,/\,\num{1010} & {\color{red}TO} & \cellcolor{blue!10}\num{41.60} & 815.79 & {\color{red}TO} & 110.05 & 471.30 & 757.15 \\
    \midrule
    \multirow{2}{=}{Embedded CPU core} & zipcpu\_busdelay\_p02 & \num{950}\,/\,\num{658} & {\color{red}TO} & 1872.35 & 1279.06 & {\color{red}TO} & \cellcolor{blue!10}\num{1243.78} & 2806.50 & {\color{red}TO} \\
    & zipcpu\_pfcache\_p16 & \num{1818}\,/\,\num{3206} & {\color{red}TO} & \cellcolor{blue!10}\num{3.27} & 1510.29 & {\color{red}TO} & 359.62 & 19.20 & 2404.34 \\
    \bottomrule
  \end{tabular*}

  \begin{tablenotes}[flushleft]\raggedright
    \scriptsize
    \item[a] \textbf{TO}: solver hit the time-out limit ($\ge$ 1 h).
    \item[b] \textbf{Pono}: results from the Pono model checker (IC3bits engine).
    \item[c] \textbf{AVR}: best of SA and SA+UF.
    \item[d] \textbf{Ablation}: results from the Pono model checker (IC3ng engine) without LLM assistance.
    \item[e] \textbf{Size}: LoC / state bits. \textbf{LoC}: number of BTOR model records; \textbf{State bits}: total bits in all BTOR state declarations.
  \end{tablenotes}
  \end{threeparttable}
  
\end{table*}

The arithmetic suite contains 33 micro-benchmarks whose safety properties depend on word-level relationships among datapath signals. These instances are intentionally small, but difficult for engines that must rediscover arithmetic invariants through bit-level generalization. They therefore provide a focused test of whether LLM-generated helper artifacts can expose the equalities, bounds, and guarded relations that PDR needs for compact inductive invariants.

\Cref{fig-simple-arith} summarizes the result for this set of problems. Pono and AVR solve 13 and 11 cases, respectively, while DreamMiner solves 19. All three \texttt{LLM4PDR} variants improve on these baselines: \texttt{LLM2Predicate} and \texttt{LLM2Clause} each solve 23 cases, and \texttt{LLM2Assert} solves 28. The cactus plot shows the same pattern in runtime. \texttt{LLM2Assert} solves the most instances and has the lowest curve over most of the solved range, while \texttt{LLM2Predicate} and \texttt{LLM2Clause} remain consistently below the deductive baselines. The arithmetic suite thus confirms that each helper-artifact interface can strengthen word-level PDR, with assertion-guided search providing the largest gain when the generated helper facts align closely with the target property.

\subsection{HWMCC Benchmarks}
To test larger designs, we selected twelve HWMCC instances~\cite{hwmcc24} spanning accelerators, embedded CPU cores, a DSP kernel, peripheral/storage controllers, and a video interface. This mix stresses both arithmetic datapaths and control-dominated state machines. Table~\ref{tab:hw-runtime} reports the wall-clock runtime.

The accelerator family \texttt{marlann\_cp\_*} shows the benefit of injecting helper artifacts into an already strong PDR engine. \texttt{LLM2Assert} is fastest on two of the three accelerator instances, and \texttt{LLM2Clause} is fastest on the third; both variants consistently improve over Pono, DreamMiner, and the IC3ng ablation. The DSP instance \texttt{intersymbol\_conv} is more decisive: Pono, AVR, and DreamMiner time out, the IC3ng ablation takes 613.64\,s, and \texttt{LLM2Predicate} proves the property in 3.06\,s. This case indicates that helper predicates can supply a compact word-level explanation that the unguided engine does not discover quickly.

The HWMCC results are heterogeneous, as expected for competition benchmarks that mix accelerators, CPU cores, DSP kernels, storage controllers, and video interfaces, and therefore LLM guidance is not uniformly best on every small instance; for example, AVR remains strong on several \texttt{picorv32} cases. However, the guided configurations provide large gains on several harder instances. On \texttt{intersymbol\_conv}, Pono, AVR, and DreamMiner time out, the unguided IC3ng ablation takes 613.64\,s, while \texttt{LLM2Predicate} proves the property in 3.06\,s. \texttt{LLM2Assert} and \texttt{LLM2Clause} also lead on different \texttt{qspiflash} and \texttt{zipcpu} cases. These results indicate that the proposed helper artifacts are complementary: predicates, clauses, and assertions each benefit different structural classes rather than acting as a single universally dominant heuristic.

\subsection{HLS--SEC Safety Kernels}
The HLS--SEC benchmarks are fixed-point Kalman-filter kernels whose top-level wrappers assert equivalence between two independently pipelined implementations. Their multipliers, adders, shifts, and long pipeline dependencies make them a stress test for arithmetic-aware verification.

\begin{table}[t]
  \centering
  \scriptsize
  \begin{threeparttable}
    \caption{HLS--SEC benchmark sizes and runtimes in seconds.\tnote{a}}
    \label{tab:hls-sec-runtime}
    \renewcommand{\arraystretch}{1.12}
    \setlength{\tabcolsep}{0.5pt}
    \begin{tabular*}{\columnwidth}{@{\extracolsep{\fill}}l c c c c c c c c@{}}
      \toprule
      \textbf{Case} & \textbf{Size}\tnote{c} & \textbf{Pono} & \textbf{AVR} &
      \textbf{Abl.} & \textbf{DM} &
      \textbf{Pred.}\tnote{b} & \textbf{Assert}\tnote{b} & \textbf{Clause}\tnote{b} \\
      \midrule
      K1 & 506/84 & 24.90 & \cellcolor{blue!10}1.49 &
        13.35 & 3.33 & 18.41 & 24.63 & 1.97 \\
      K2 & 491/102 & 93.57 &
        998.46 &
        1217.89 &
        \cellcolor{blue!10}6.47 &
        20.45 & 24.64 & 25.01 \\
      K3 & \multirow{2}{*}{424/83} & \textcolor{red}{TO} &
        \textcolor{red}{TO} &
        \textcolor{red}{TO} &
        5275.60 &
        9912.86 &
        5112.58 &
        \cellcolor{blue!10}3990.83 \\
      K4 &  & \textcolor{red}{TO} &
        \textcolor{red}{TO} &
        \textcolor{red}{TO} &
        5366.69 &
        9741.71 &
        5184.17 &
        \cellcolor{blue!10}3664.56 \\
      \bottomrule
    \end{tabular*}

    \begin{tablenotes}[flushleft]\raggedright
      \scriptsize
      \item[a] \textbf{K1--K4}: Kalman1--4. \textbf{TO}: $\ge$ 3 h. \textbf{Abl.}: IC3ng without LLM assistance. \textbf{DM}: DreamMiner.
      \item[b] \textbf{Pred./Assert/Clause}: LLM2Predicate, LLM2Assert, and LLM2Clause.
      \item[c] \textbf{Size}: RTL LoC / state bits.
    \end{tablenotes}
  \end{threeparttable}
\end{table}

\begin{table*}[t]
  \centering
  \scriptsize
  \begin{threeparttable}
    \caption{Benchmark sizes and runtime comparison on open-source circuit designs with assertions. Runtimes are in seconds.\tnote{a}}
    \label{tab:opensrc-circuit-runtime}
    {
    \renewcommand{\arraystretch}{1.12}
    \setlength{\tabcolsep}{1.5pt}
    \sisetup{
      table-format         = 3.2,
      group-digits         = false,
      detect-weight        = true,
      detect-inline-weight = math
    }

    \begin{tabular*}{\linewidth}{@{\extracolsep{\fill}}>{\raggedright\arraybackslash}p{2.0cm} l c S S S S S S S@{}}
      \toprule
      \textbf{Category} & \textbf{Benchmark} &
      \textbf{Size}\tnote{b} & \textbf{Pono} & \textbf{AVR} &
      \textbf{Ablation} & \textbf{dreamminer} &
      {\shortstack{\texttt{LLM2}\\\texttt{Predicate}}} & {\shortstack{\texttt{LLM2}\\\texttt{Assert}}} & {\shortstack{\texttt{LLM2}\\\texttt{Clause}}} \\
      \midrule
      \multirow{3}{=}{Bus interconnects, bridges, etc.} & axil2axis & \num{1043}\,/\,\num{1802} & 79.18 & \multicolumn{1}{c}{\textcolor{red}{TO}} & 32.01 & 80.43 & \cellcolor{blue!10}{\num{25.37}} & 69.80 & 29.43 \\
      & wbm2axilite & \num{391}\,/\,\num{521} & \multicolumn{1}{c}{\textcolor{red}{TO}} & \cellcolor{blue!10}{\num{2.58}} & 47.01 & \multicolumn{1}{c}{\textcolor{red}{TO}} & 25.71 & \multicolumn{1}{c}{\textcolor{red}{TO}} & 36.15 \\
      & axildouble & \num{1257}\,/\,\num{2448} & 149.25 & \multicolumn{1}{c}{\textcolor{red}{TO}} & 253.19 & 155.44 & \cellcolor{blue!10}{\num{131.60}} & 161.20 & 193.30 \\
      \midrule
      \multirow{3}{=}{DDR3 Controller} & multiconfig\_ecc & \multirow{3}{*}{\num{4136}\,/\,\num{10130}} & 436.71 & \multicolumn{1}{c}{\textcolor{red}{TO}} & 293.76 & 450.28 & \cellcolor{blue!10}{\num{147.08}} & 300.60 & 237.89 \\
      & singleconfig &  & 443.23 & \multicolumn{1}{c}{\textcolor{red}{TO}} & 313.21 & 441.98 & \cellcolor{blue!10}{\num{153.03}} & 328.69 & 255.23 \\
      & multiconfig\_default &  & 600.87 & \multicolumn{1}{c}{\textcolor{red}{TO}} & 418.60 & 597.62 & \cellcolor{blue!10}{\num{124.88}} & 475.10 & 139.70 \\
      \bottomrule
    \end{tabular*}}

    \begin{tablenotes}[flushleft]\raggedright
      \scriptsize
      \item[a] \textbf{TO}: solver hit the time-out limit ($\ge$ 1 h).
      \item[b] \textbf{Size}: LoC / state bits. \textbf{LoC}: RTL source code lines; \textbf{State bits}: total bits in all BTOR state declarations.
    \end{tablenotes}
  \end{threeparttable}
  
\end{table*}

Table~\ref{tab:hls-sec-runtime} shows a clear progression in difficulty. On \texttt{Kalman1}, AVR is fastest and \texttt{LLM2Clause} is close behind. On \texttt{Kalman2}, DreamMiner achieves the best runtime, while \texttt{LLM2Predicate} remains much faster than Pono, AVR, and the ablation. The harder \texttt{Kalman3} and \texttt{Kalman4} cases separate the approaches more clearly: Pono, AVR, and the ablation time out, while the \texttt{LLM4PDR} variants complete the proofs. Among our variants, \texttt{LLM2Clause} is fastest on both hard kernels, completing them in about one hour. These results suggest that side-loaded clauses are particularly effective for deep HLS pipelines, where useful arithmetic relations can be validated early and propagated through the PDR frame sequence.

\subsection{Open-Source RTL Designs with Assertions}
The open-source suite in Table~\ref{tab:opensrc-circuit-runtime} contains two families of real-world RTL designs. The WB2AXIP cases include fully pipelined AXI4 and AXI-Lite bridges that allow multiple in-flight transactions, creating deep handshake logic. The DDR3 controller cases add rank-bank management, write levelling, read calibration, and optional SEC--DED error correction, yielding long control pipelines and wide datapaths. These designs are formally annotated, and their properties are challenging for general-purpose model checkers.

\texttt{LLM2Predicate} is the strongest configuration on this suite. It obtains the best runtime on five out of six cases and remains competitive on \texttt{wbm2axilite}, where AVR is exceptionally fast. On the AXI bridge cases, \texttt{LLM2Predicate} outperforms Pono, the ablation, DreamMiner, and the other \texttt{LLM4PDR} modes on \texttt{axil2axis} and \texttt{axildouble}; it also solves \texttt{wbm2axilite} while Pono and DreamMiner time out. On the DDR3 controllers, AVR times out on all three cases, whereas \texttt{LLM2Predicate} gives roughly $3$--$5\times$ speedups over Pono and clear improvements over the IC3ng ablation. This pattern is consistent with the nature of these designs: predicate guidance is well suited to repeated control/data relationships, such as handshake consistency and pipeline-state correspondence, that become useful during many MIC calls.

\subsection{Ablation: IC3bits vs.\ IC3ng}
\begin{figure}[t]
\centering
\includegraphics[width=0.82\linewidth]{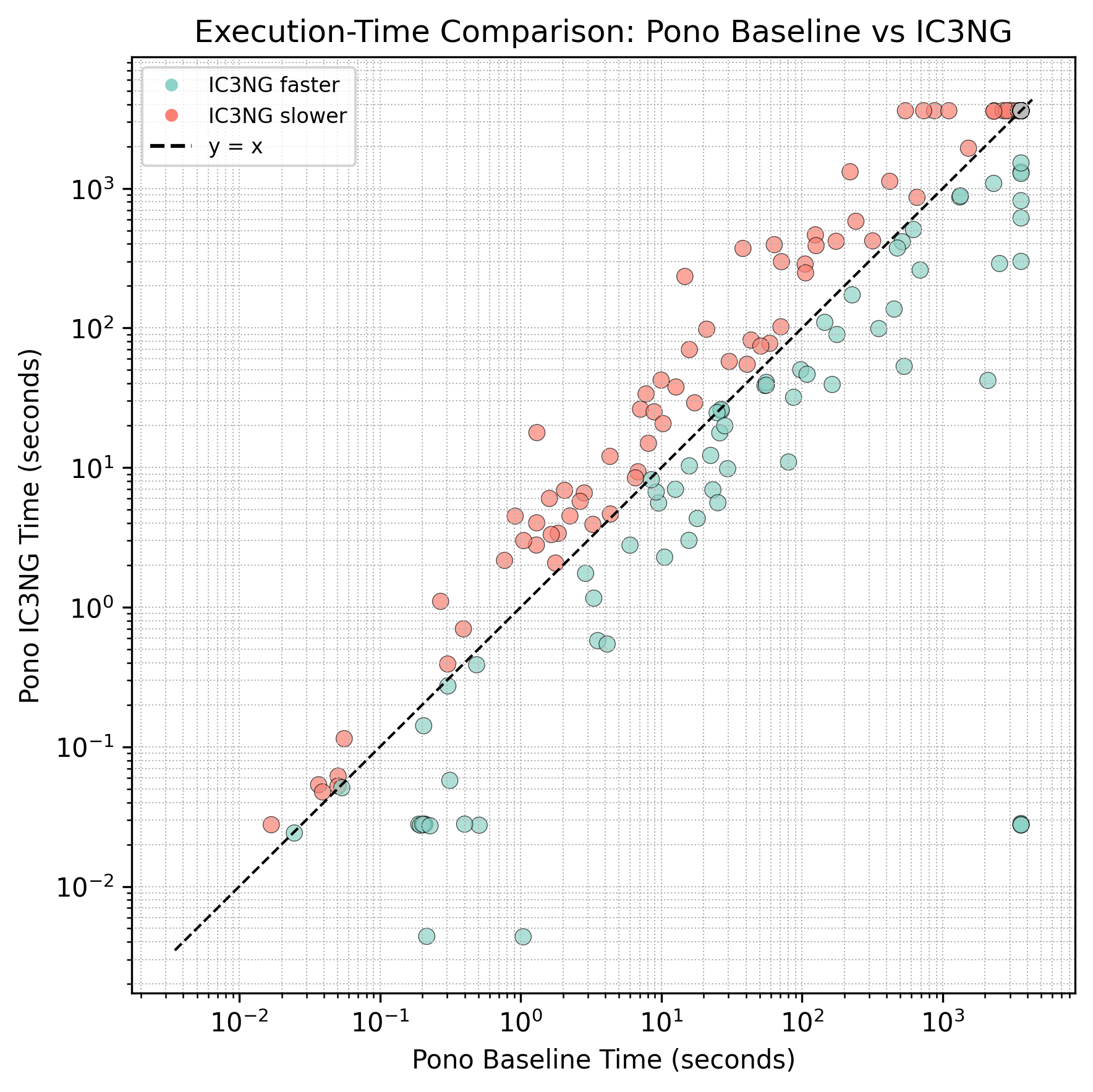}
\caption{Execution-time scatter plot comparing Pono IC3bits with Pono IC3ng without helper artifacts.}
\label{fig-llm4pdr-ablation}
\end{figure}

To better exploit LLM-generated helpers, we use IC3ng, which extends IC3bits with the \textsc{Down}/MIC-style inductive generalization~\cite{bradley2011ic3} and eager lemma propagation during blocking, following the rationale of aggressive clause pushing~\cite{ivrii2015pushing}. These mechanisms support helper-guided proof search: the generalizer is extended to retain word-level helpers while reducing bit-level literals, and eager propagation makes the resulting lemmas available sooner in subsequent frames. To establish that our speedups cannot be explained by these engine changes alone, we disable all LLM-generated helpers while retaining the underlying IC3ng heuristics. \Cref{fig-llm4pdr-ablation} compares this unguided configuration with IC3bits; the Ablation columns in the tables compare it with \texttt{LLM4PDR} on the same engine.

Without helpers, IC3ng does not consistently outperform IC3bits: \Cref{fig-llm4pdr-ablation} shows both improvements and regressions. Thus, the added heuristics alone do not establish a general performance advantage. Enabling helpers on the same IC3ng engine yields substantial additional speedups on several benchmarks; for example, \texttt{LLM2Predicate} reduces \texttt{intersymbol\_conv} from 613.64\,s to 3.06\,s. These results show that the engine changes alone cannot explain these gains: the LLM-generated semantic information makes an additional contribution when used together with the enhanced generalization and propagation procedures.

\section{Conclusion}
\label{sec:conclusion}
This paper presents \texttt{LLM4PDR}, a framework for injecting LLM-generated semantic guidance into PDR-based hardware model checking. The key idea is to keep the LLM outside the soundness boundary: it proposes predicates, clauses, and helper assertions, while the verifier parses, type-checks, validates, and refines those artifacts through ordinary PDR proof obligations. This lets the engine benefit from high-level word-level relations without trusting the LLM as an oracle.

Across arithmetic, HWMCC, HLS--SEC, and open-source RTL benchmarks, the three \texttt{LLM4PDR} modes improve different parts of the proof search. Assertion guidance is strongest on the arithmetic suite, clause guidance is especially effective on deep HLS pipelines, and predicate guidance performs best on open-source control-plus-datapath designs. The ablation against unguided IC3ng further shows that these gains come from the generated helper artifacts rather than from an engine swap alone. These results indicate that LLM-generated semantic hints can be a practical and sound complement to conventional PDR generalization.


\section*{Acknowledgments}
The authors used Gemini~3.8 Flash only to improve the language and readability of the manuscript text. It was not used to generate technical ideas, results and figures. All suggested revisions were reviewed and verified by the authors, who take full responsibility for the final content.

\ifdefined\IEEEFORMAT
  \nobalance
\else
  \balance
\fi
\nocite{*}
\bibliography{refs}

\ifdefined\IEEEFORMAT
\makeatletter
\def\@IEEEBIOskipN{12pt}
\xpatchcmd{\IEEEbiography}
  {\vskip \@IEEEBIOskipN plus 1fil minus 0\baselineskip}
  {\vskip \@IEEEBIOskipN}
  {}{\PackageError{biographies}{Unable to set biography spacing}{Check the IEEEtran biography definition.}}
\makeatother


\begin{IEEEbiography}[{\includegraphics[width=1in,height=1.25in,clip,keepaspectratio]{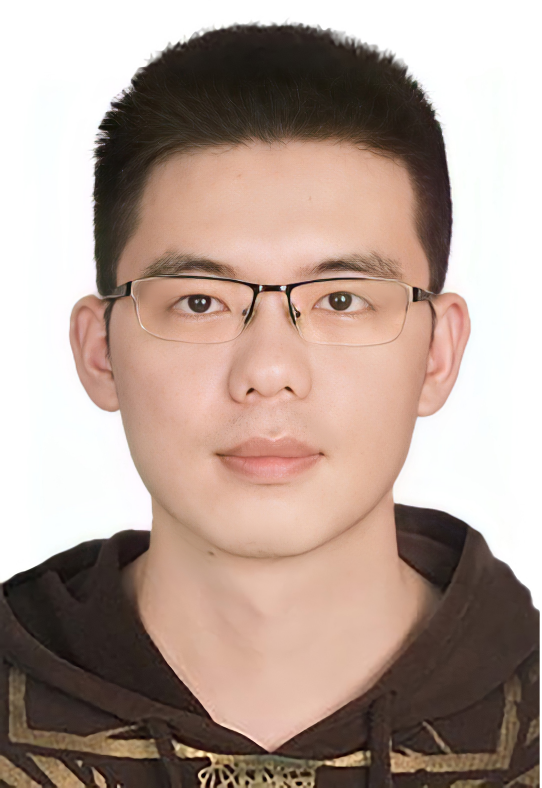}}]{Guangyu Hu}
received the B.E. degree in software engineering from Xiamen University, Xiamen, China, in 2021.

He received his PhD degree in Microelectronics, Individualized Interdisciplinary Program (IIP) of Hong Kong University of Science and Technology, Clear Water Bay, Hong Kong SAR. His research interests include hardware model checking and logic synthesis.

\end{IEEEbiography}

\begin{IEEEbiography}[{\includegraphics[width=1in,height=1.25in,clip,keepaspectratio]{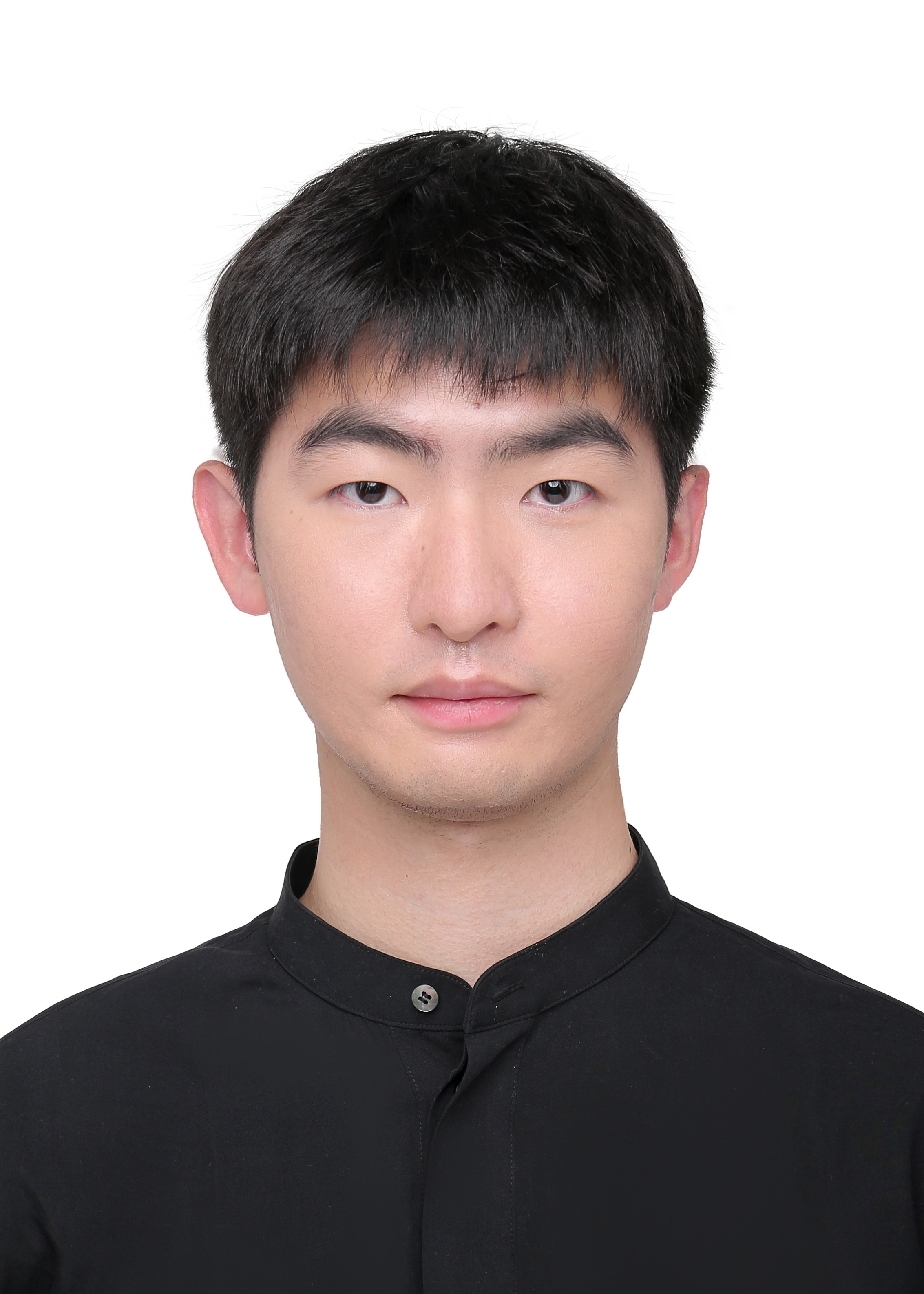}}]{Mingkai Miao}
received the B.Eng. degree in electronic and information engineering from the University of Electronic Science and Technology of China, Chengdu, China, in 2021, and the M.Eng. degree in electrical and computer engineering from the University of Illinois Urbana-Champaign, Urbana, IL, USA, in 2024. He is currently pursuing the Ph.D. degree in microelectronics with the Hong Kong University of Science and Technology (Guangzhou), Guangzhou, China. His research interests include hardware formal verification, IC3/PDR model checking, and machine learning and large language model techniques for electronic design automation.

\end{IEEEbiography}

\begin{IEEEbiography}[{\includegraphics[width=1in,height=1.25in,clip,keepaspectratio]{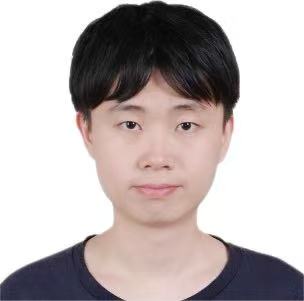}}]{Zhiyuan Yan}
received the B.Sc. degree in Electronic Information Science and Technology from Henan University, Kaifeng, China, in 2020, and the master's degree in Microelectronics from Nanyang Technological University, Singapore, in 2022. He is currently pursuing the Ph.D. degree with the Microelectronics Thrust, Hong Kong University of Science and Technology (Guangzhou), Guangzhou, China, advised by Prof. Hongce Zhang. His research lies in AI for EDA, especially machine learning-guided formal verification and SAT.

\end{IEEEbiography}

\begin{IEEEbiography}[{\includegraphics[width=1in,height=1.25in,clip,keepaspectratio]{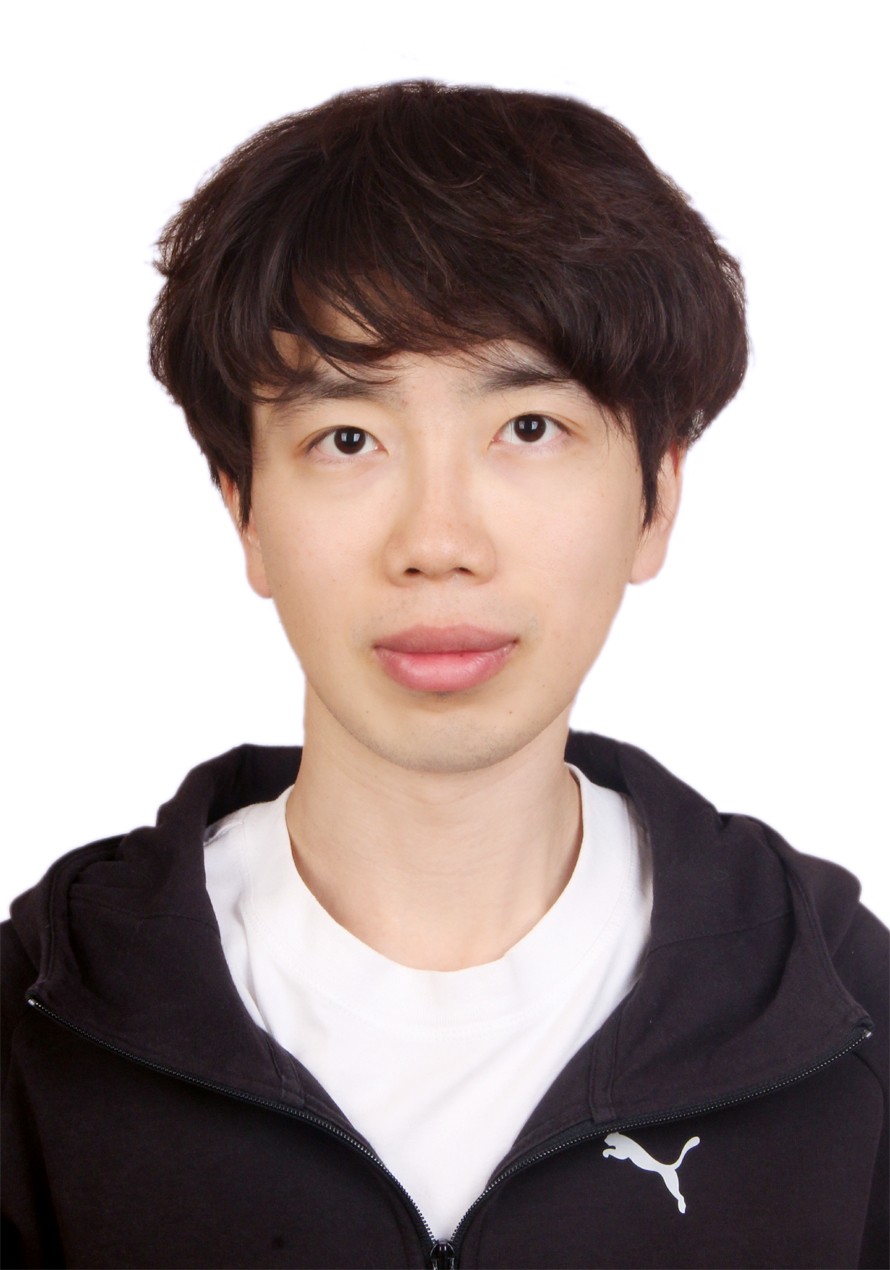}}]{Xiaofeng Zhou}
received the B.Eng. degree in information science and engineering from Southeast University, Nanjing, China, in 2020, and the Ph.D. degree in electronic and computer engineering from The Hong Kong University of Science and Technology, Hong Kong, in 2026. His research interests include hardware formal verification, model checking, high-level synthesis, and FPGA-based acceleration.

\end{IEEEbiography}

\begin{IEEEbiography}[{\includegraphics[width=1in,height=1.25in,clip,keepaspectratio]{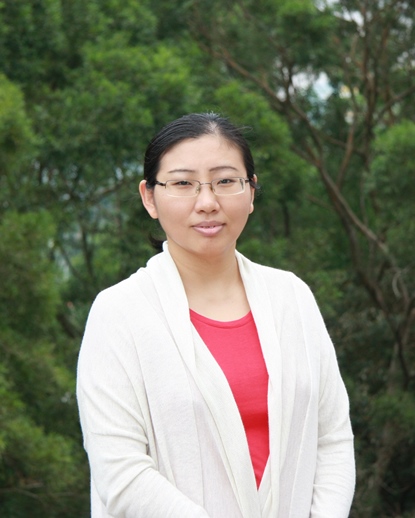}}]{Wei Zhang}
(Fellow, IEEE) received the Ph.D. degree from Princeton University, Princeton, NJ, USA, in 2009.

She was an Assistant Professor with the School of Computer Engineering, Nanyang Technological University, Singapore, from 2010 to 2013. She joined The Hong Kong University of Science and Technology, Hong Kong, in 2013, where she is currently a Professor and she established the Reconfigurable Computing System Laboratory. Her research interests include reconfigurable systems, field-programmable gate array-based design, low-power high-performance multicore systems, electronic design automation, embedded systems, and emerging technologies.
\end{IEEEbiography}

\begin{IEEEbiography}[{\includegraphics[width=1in,height=1.25in,clip,keepaspectratio]{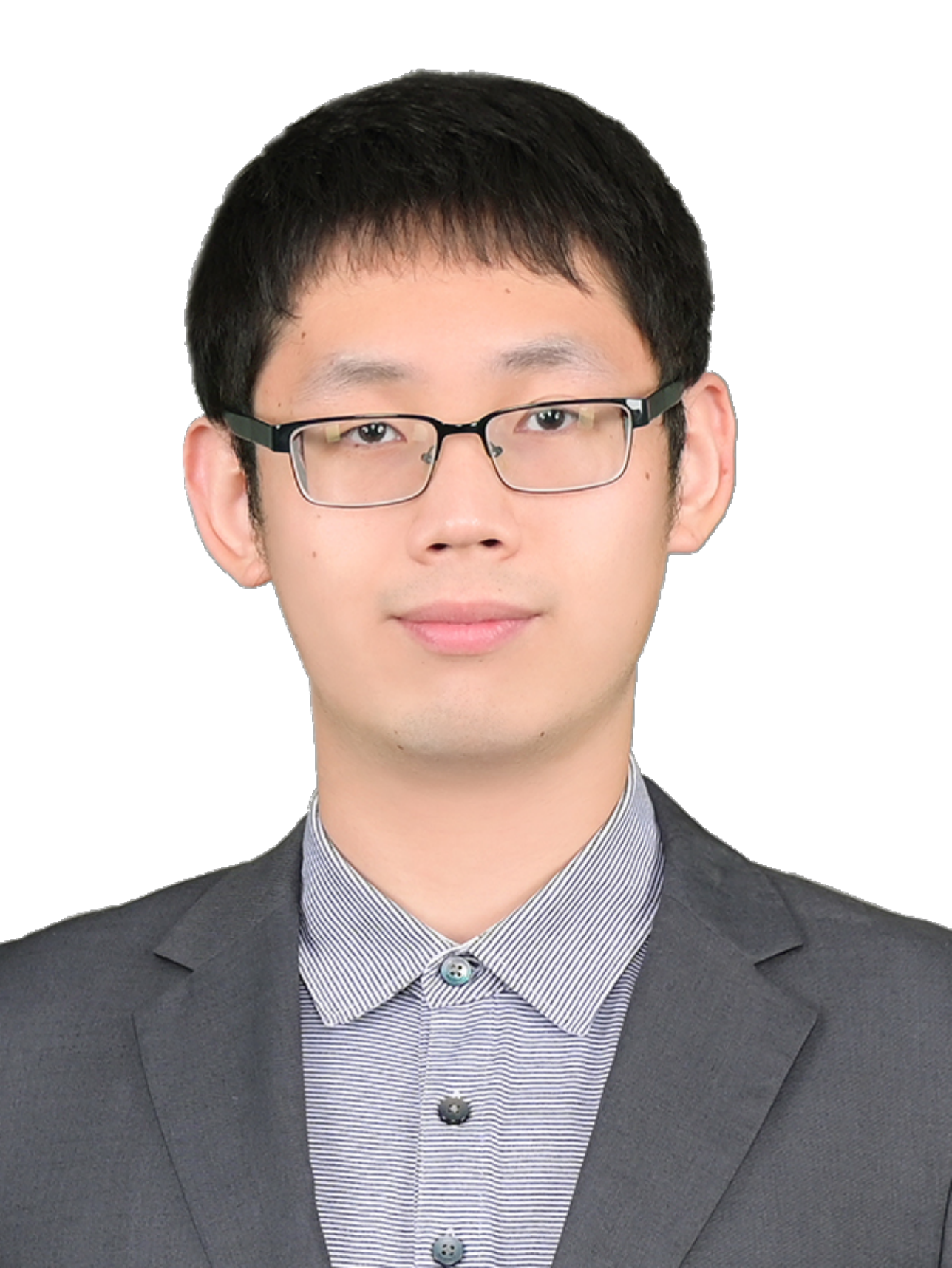}}]{Hongce Zhang}
(Member, IEEE) received the B.S. degree in microelectronics from Shanghai Jiao Tong University, Shanghai, China, in 2015, and the Ph.D. degree from the Electrical and Computer Engineering Department of Princeton University, NJ, USA, in 2021.

He is currently an Assistant Professor with the Microelectronics Thrust, Function Hub of Hong Kong University of Science and Technology (Guangzhou), Guangzhou, China, and is also affiliated with the Division of Emerging Interdisciplinary Areas (EMIA) of the Hong Kong University of Science and Technology, Clear Water Bay, Hong Kong SAR. His research interests include formal verification and hardware model checking.

\end{IEEEbiography}

\fi

\end{document}